\documentclass[preprint2]{aastex}

\shortauthors{Vavrycuk}

\usepackage{graphicx}
\usepackage{amsmath}
\usepackage{esint}

\usepackage{natbib}

\begin{document} 

\title{Universe opacity and cosmic dynamics}

\author{V. Vavry\v cuk}

\affil{The Czech Academy of Sciences}
\affil{Bo\v cn\' i II 1401, 141 00 Praha 4}
\email{vv@ig.cas.cz}

\received{December 27, 2018}

\begin{abstract}
Equations of cosmic dynamics for a model of the opaque Universe are derived and tested on supernovae (SNe) Ia observations. The model predicts a cyclic expansion/contraction evolution of the Universe within a limited range of scale factors. The maximum scale factor is controlled by the overcritical density of the Universe, the minimum scale factor depends on global stellar and dust masses in the Universe. During contraction due to gravitational forces, the extragalactic background light (EBL) and intergalactic opacity increase with time because of a smaller proper volume of the Universe. The radiation pressure produced by absorption of the EBL by dust steeply rises and counterbalances the gravitational forces. The maximum redshift, at which the radiation pressure is capable to stop the contraction and start a new expansion, is between 15 and 45. The model avoids the Big Bang and concepts of the non-baryonic dark matter and dark energy. The model successfully explains the existence of very old mature galaxies at high redshifts and observed dimming of the luminosity of SNe Ia with redshift. The cosmic microwave background (CMB) is interpreted as thermal radiation of intergalactic dust and its temperature is predicted with accuracy higher than 2\%. The CMB temperature anisotropies are caused by the EBL fluctuations related to large-scale structures in the Universe. A strong decline of the luminosity density at $z>4$ is explained by high opacity of the Universe rather than by its darkness due to a missing stellar mass at high redshifts.
\end{abstract}


\section{Introduction}

The Big Bang theory is supported by the three key observations: (1) the expansion of the Universe measured by the redshift of distant galaxies, (2) the existence of the cosmic microwave background (CMB), and (3) the relative amounts of hydrogen, helium and deuterium in the Universe. The expansion of the Universe was predicted by Friedmann in 1922 \citep{Friedmann1999} and confirmed experimentally by \citet{Lemaitre1927} and \citet{Hubble1929}. The expansion velocity of $\sim 500\, \mathrm{km\,s^{-1}\, Mpc^{-1}}$ estimated originally by Hubble was later corrected to $\sim 67-73\, \mathrm{km\,s^{-1}\, Mpc^{-1}}$ obtained, e.g., by the baryonic acoustic peaks in the CMB spectrum \citep{Ade2014b} or by a distance ladder method based on measurements of SN Ia supernovae in galaxies at distance of hundreds of Mpc \citep{Freedman2001, Riess2011, Riess2016, Riess2018}. The CMB as a relic radiation of the Big Bang has been predicted by several physicists and cosmologists \citep{Alpher1948, Gamow1952, Gamow1956}, and experimentally detected by \citet{Penzias1965}. Later accurate CMB observations proved that the CMB has almost a perfect blackbody spectrum with an average temperature of $T = 2.7 K$ \citep{Fixsen2009} and with small-scale fluctuations of $\pm 70 \, \mu\mathrm{K}$ \citep{Bennett2003, Hinshaw2009, Bennett2013}. Interpreting acoustic peaks in the CMB spectrum caused by these fluctuations led to estimation of the curvature of the universe and the dark matter density \citep{Hu_Dodelson2002, Spergel2007, Komatsu2011}. Finally, the Big Bang nucleosynthesis (BBN) theory predicted the primordial abundances of deuterium, helium and lithium \citep{Olive2000, Cyburt2016}. The predicted abundance of helium $^{4}\mathrm{He/H} = 0.2470$ was confirmed by observations and deuterium abundance was used for estimation of the baryon density in the Universe. 

Although it seems that the validity of the Big Bang theory is apparently undisputable and verified by many independent observations, a deeper insight reveals essential difficulties and controversies of this theory. First, the universe expansion itself cannot be taken as evidence of the Big Bang. The observations of the expansion are limited to redshifts less than $\sim 14-15$ and any inferences about the cosmic dynamics at redshifts above this limit should be taken with caution. Obviously, extrapolations back to a primordial singularity are highly speculative. 

Second, it is assumed that the CMB provides information about the universe at very high redshifts, but the origin of the CMB as a relic radiation of the Big Bang is, in fact, uncertain. For example, accurate measurements of the CMB anisotropies by WMAP and Planck revealed several unexpected features at large angular scales such as non-Gaussianity of the CMB \citep{Vielva2004, Cruz2005, Planck_XXIV_2014} and a violation of statistical isotropy and scale invariance following from the Big Bang theory \citep{Schwarz2016}. Moreover, \citet{Vavrycuk2017b} argues that the CMB cannot be a relic radiation because the blackbody spectrum of the relic radiation produced at very high redshifts ($z \sim 1100$) should be distorted by galactic and intergalactic dust in the lower-redshift Universe ($z < 10$). The distortion should be well above the sensitivity of the COBE/FIRAS, WMAP, or Planck flux measurements \citep{Fixsen1996, Hinshaw2009, Planck2013_Results} and should include a decline of the spectral as well as total CMB intensity due to absorption. Obviously, if the CMB is not a relic radiation, the concept of the Big Bang is seriously disputed.

Third, also evidence of the Big Bang provided by the BBN is questionable. In order to fit observations of deuterium abundance, the baryon-to-photon ratio η or equivalently the baryon density $\Omega_b$ is determined. Since this value is many times lower than the mass density of the Universe indicated from gravitational measurements, a concept of the non-baryonic dark matter had to be introduced. In addition, the predictions of the $^{4}$He, and $^{7}$Li abundances by the BBN are not very persuasive. Initially, observations of $^{4}$He/H abundance did not match the predicted value well \citep{Pagel1992, Peimbert2000} but after two decades of efforts \citep{Peimbert2007, Izotov2014, Aver2015} when adopting a large number of systematic and statistical corrections \citep[their table 7]{Peimbert2007}, a satisfactory fit has been achieved \citep[his fig. 14]{Vavrycuk2018}. By contrast, the predicted $^{7}$Li/H abundance is $2 - 3$ times larger than observations \citep{Cyburt2008, Fields2011} and to date, there is no solution of the discrepancy of the $^{7}$Li abundance without substantial departures of the BBN theory \citep{Cyburt2016}.

The Big Bang theory and the interpretation of the CMB as a relic radiation of the Big Bang have also some other unsolved problems. It is deduced from supernovae measurements and from the CMB anisotropy that the universe is nearly flat at present \citep{Ade2014b, Planck_2015_XIII, Planck_2018_VI}. This implies that the deviation of the universe curvature from zero was less than $10^{-10}$ at the time of the Big Bang nucleosynthesis and the actual density of the Universe was almost exactly equal to the critical density, which is unlikely (the 'flatness problem', see \citet{Ryden2016}). Similarly, since the Universe is nearly homogeneous and isotropic on large scale at present, it should be almost exactly isotropic and homogeneous at the last scattering epoch, even though that the matter at different parts of the space was not in a causal contact (the 'horizon problem', see \citet{Ryden2016}). Furthermore, except for considering the non-baryonic dark matter, a 'dark energy' concept as energy of vacuum was introduced in order to interpret the supernovae data in terms of an accelerating expansion of the Universe at $z \sim 0.5$ \citep{Riess1998, Perlmutter1999, Li2013}. However, it is unclear why the dark energy and the matter energy density have the same order of magnitude just at the present epoch, although the matter density changed by a factor of $10^{42}$ during the universe evolution (the 'coincidence problem', see \citet{Weinberg2013, Bull2016}). Finally, recent observations of very old galaxies at redshifts $z \sim 8$ or higher question the age of the Universe predicted by the Big Bang theory \citep{Ellis2013, Oesch2016, Hashimoto2018}.
	
Many of the mentioned problems of the Big Bang theory are readily solved under an alternative cosmological model based on the idea that the CMB is not relic radiation of the Big Bang but thermal radiation of intergalactic dust \citep{Wright1982, Wright1987, Aguirre2000}. This model was recently revived and revisited by \citet{Vavrycuk2018} who suggested that a virtually transparent Universe at the present epoch was increasingly opaque at higher redshifts. The temperature of intergalactic dust increases as $(1+z)$ in this model and compensates exactly the change of wavelengths of dust radiation caused by redshift. The theory predicts the CMB temperature with the accuracy higher than 2\%. The CMB temperature fluctuations are caused by fluctuations of the extragalactic background light (EBL) produced by galaxy clusters and voids in the Universe. The CMB polarization anomalies originate in the polarized thermal emission of needle-shaped conducting dust grains which are aligned by magnetic fields around large-scale structures such as clusters and voids. The number density of galaxies and the overall dust masses within galaxies and in the intergalactic space are basically constant with cosmic time. A strong decline of the luminosity density for $z > 4$ is interpreted as the result of high opacity of the high-redshift universe rather than of a decline of the global stellar mass density at high redshifts. Instead of a Universe evolution and expansion from an initial singularity, an idea of steady-state universe oscillations within a limited range of scale factors is suggested \citep{Vavrycuk2018}. 

Although the interpretation of the CMB as thermal radiation of dust is well developed theoretically and also supported by observations \citep{Vavrycuk2018}, equations of cosmic dynamics under the cyclic universe model are missing. In particular, it is unclear which forces act against the gravity being able to cause a periodic expansion/contraction evolution of the Universe. This relates to the old problem addressed originally by \citet{Einstein1917} who introduced a cosmological constant into his field equations of general relativity to counterbalance the gravity in the static universe. A similar idea has been adopted when the dark energy was introduced to explain an accelerating expansion reported by teams conducting and interpreting supernovae measurements \citep{Riess1998, Perlmutter1999}. In this paper, I show that the mysterious dark energy concept is not needed and that the repulsive forces in the Universe might be of a standard physical origin. I show that the key to resolving this fundamental long-lasting cosmological puzzle is the opacity of the Universe and processes related to absorption of light in the Universe. I document that the universe opacity, a central idea of the dust theory \citep{Vavrycuk2016, Vavrycuk2017a}, offers also a simple and physically understandable solution even for the cosmic dynamics. I derive formulas for radiation pressure produced by absorption of light by galaxies, Lyman α systems and by intergalactic dust. I demonstrate that the light pressure is negligible at the present epoch, but it could be significantly stronger in the past epochs and it might balance the gravitational forces at high redshifts. Based on the numerical modeling and observations of basic cosmological parameters, I estimate the maximum redshift of the Universe achieved in the past and the maximum scale factor of the Universe which can be achieved in future. Finally, the predicted cosmic dynamics is confronted with current supernovae observations.

\section{Theory}

\subsection{Equation of motion for two galaxies}

\begin{figure*}
\includegraphics[angle=0,width=13cm,trim=50 160 210 80]{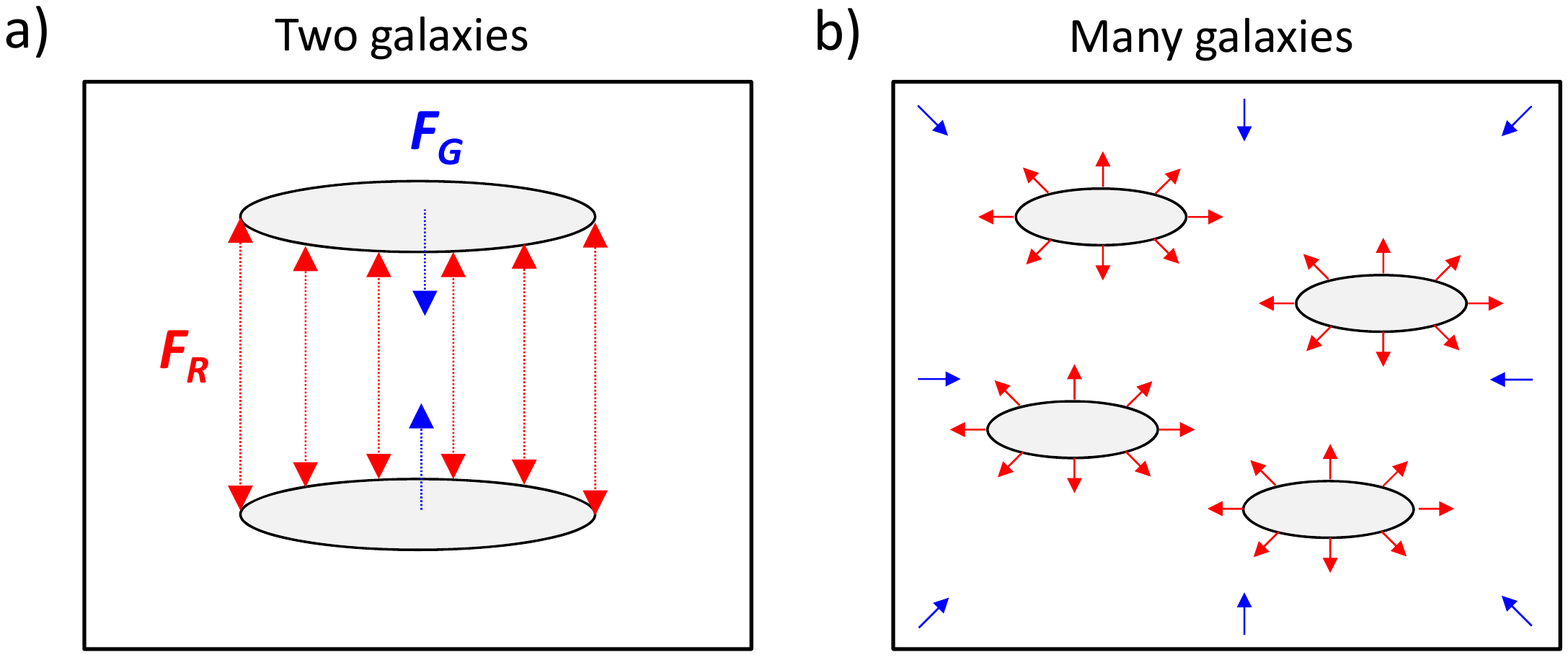}
\caption{
Gravitational forces and radiation pressure acting on galaxies. Blue - compressional gravitational forces, red - repulsive radiation pressure, gray - galactic dust absorbing light.
}
\label{fig:1}
\end{figure*}

Let us assume two galaxies interacting mutually by the Newtonian gravity and by light radiation
\begin{equation}\label{eq1}
M\ddot R = -\frac{GM^{2}}{R^{2}} + Sp \,,
\end{equation}
where $M$ is the mass of a galaxy, $R$ is the distance between the galaxies, $S$ is the cross-section of a galaxy, $p$ is the pressure caused by light radiated by one galaxy and absorbed by galactic dust in the other galaxy, and $G$ is the gravitational constant. The gravity and radiation forces have opposite signs, because they act against each other (see Fig.~\ref{fig:1}a) . The radiation pressure produced by the galaxies reads
\begin{equation}\label{eq2}
p = \frac{\kappa}{c}\frac{L}{4\pi R^{2}} \,,
\end{equation}
where $\kappa$ is the galactic opacity, $c$ is the speed of light, and $L$ is the galactic luminosity (in W). Hence the ratio between the gravitational and radiation forces 
\begin{equation}\label{eq3}
\frac{F_G}{F_R} = 4 \pi c G \frac{M}{L}\frac{M}{\kappa S} 
\end{equation}
is independent of the distance $R$ between galaxies. Considering the galaxy mass $M = 6 \times 10^{10} \, M_{\odot}$, mass-to-light ratio $M/L = 5 M_{\odot}/L_{\odot}$, galactic opacity $\kappa = 0.25$ and a spherical shape of a galaxy with radius $r = 10$ kpc, we get for the ratio $F_G/F_R$ a value of about $1 \times 10^{4}$.

If we consider two stars with parameters of the Sun instead of two galaxies and with the star opacity $\kappa = 1$, the ratio $F_G/F_R$ attains a value of $1.7 \times 10^{15}$. Hence, the gravity attraction is higher than the radiation repulsion by more than 15 orders for stars but only by 4 orders for galaxies. The ratio for galaxies may further decrease when we take into account that the radiation in intergalactic space is, in fact, the EBL and its intensity $I^{\mathrm{EBL}}$ increases with redshift as \citep[his eq. 5]{Vavrycuk2018} 
\begin{equation}\label{eq4}
I^{\mathrm{EBL}} = I^{\mathrm{EBL}}_0 \left(1+z\right)^4 \,. 
\end{equation}
Consequently, radiation pressure $p$ caused by absorption of the EBL by galactic dust 
\begin{equation}\label{eq5}
p = \frac{\kappa}{c} I^{\mathrm{EBL}}
\end{equation}
increases according to the same redshift dependence as $I^{\mathrm{EBL}}$ in eq. (4)
\begin{equation}\label{eq6}
p = p_0 \left(1+z\right)^4 \,. 
\end{equation}
Hence, the $F_G/F_R$ ratio might be significantly lower in the past epochs than at the present time.

Inserting the relative scale factor $a = R/R_0 = \left(1+z\right)^{-1}$ and eq. (4) into eq. (1) 
\begin{equation}\label{eq7}
\ddot a = -\frac{GM}{R^{3}_0} \frac{1}{a^2} + \frac{S}{M} \frac{p_0}{R_0} \frac{1}{a^4} \,, 
\end{equation}
multiplying by $\dot a$ and integrating in time, we get
\begin{equation}\label{eq8}
\frac{1}{2} {\left({\frac{\dot a}{a}}\right)}^2 = \frac{GM}{R^{3}_0} \frac{1}{a^3} - \frac{1}{3} \frac{S}{M} \frac{p_0}{R_0} \frac{1}{a^5} + \frac{U} {a^2} \,, 
\end{equation}
where the subscript '0' denotes the quantity at the present time, and $U$ is the integration constant expressing a normalized energy of the system.

\subsection{Dynamic equation of the transparent universe}

If we consider an expanding transparent universe with a uniform random distribution of galaxies (see Fig.~\ref{fig:1}b), eq. (8) must be modified. The gravitational term in eq. (8) reads:
\begin{equation}\label{eq9}
\frac{GM}{R^{3}_0} \frac{1}{a^3} = \frac{4}{3} \pi G \rho_0 \frac{1}{a^3} \,, 
\end{equation}
where $M$ means the mass of the Universe within a sphere of radius $R_0$, $\rho_0$ is the density of the Universe at the present time, and $R_0$ is the mean distance between galaxies at the present time.

The radiation-absorption term in eq. (8) reads:
\begin{equation}\label{eq10}
\frac{1}{3} \frac{S}{M} \frac{p_0}{R_0} \frac{1}{a^5} = \frac{4}{3} \frac{p_0}{\gamma_0 \rho_0 R_0} \frac{1}{a^5} \,, 
\end{equation}
where $S$ is the total surface of galaxies within a sphere of radius $R_0$, $p_0$ is the pressure acting on a galaxy surface, $\gamma_0$ is the mean-free path between galaxies, subscript '0' denotes quantities at the present time and we used the following relations
\begin{equation}\label{eq11}
\frac{S}{M} = \frac{4 \pi r^2 n V}{\rho_0 V} = \frac{4}{\gamma_0 \rho_0} \,, 
\end{equation}
\begin{equation}\label{eq12}
\gamma_0 = \frac{1}{4 \pi r^2 n}  \,, 
\end{equation}
where $n$ is the galaxy number density, and $r$ is the mean galaxy radius.

Inserting eqs. (5), (9) and (10) into eq. (8), we get
\begin{equation}\label{eq13}
{\left({\frac{\dot a}{a}}\right)}^2 = \frac{8}{3} \pi G \rho_0 \frac{1}{a^3} - \frac{8}{3} \frac{\kappa}{\gamma_0} \frac{I^{\mathrm{EBL}}_0}{c \rho_0 R_0} \frac{1}{a^5} + \frac{2 U} {a^2} \,. 
\end{equation}
%

\subsection{Dynamic equation of the opaque universe}
If we consider an intergalactic medium (IGM) with dust, the term expressing absorption of light by galactic dust in eq. (13)
\begin{equation}\label{eq14}
\lambda^G_0 = \frac{\kappa}{\gamma_0} 
\end{equation}
must be replaced by total absorption $\lambda_0$ which includes also absorption by intergalactic dust $\lambda^{IG}_0$  
\begin{equation}\label{eq15}
\lambda_0 = \lambda^G_0 + \lambda^{IG}_0 \,.
\end{equation}
Taking into account the relation between the EBL intensity $I^{\mathrm{EBL}}$ and the luminosity density $j$ \citep[his eq. 7]{Vavrycuk2018}
\begin{equation}\label{eq16}
\lambda_0 I^{\mathrm{EBL}}_0 = \frac{j_0}{4 \pi} \,,
\end{equation}
we get
\begin{equation}\label{eq17}
{\left({\frac{\dot a}{a}}\right)}^2 = \frac{8}{3} \pi G \rho_0 \frac{1}{a^3} - \frac{2}{3 \pi} \frac{j_0}{c \rho_0 R_0} \frac{1}{a^5} + \frac{2 U} {a^2} \,, 
\end{equation}
where $\rho_0$ is the mass density of the Universe, $j_0$ is the luminosity density, $U$ is the normalized energy of the system of galaxies and IGM, and $R_0$ is the mean distance between galaxies which can be calculated from the galaxy number density $n$ as follows
\begin{equation}\label{eq18}
R_0 = \left[{\frac{3}{4 \pi n}}\right]^{\frac{1}{3}} \,.
\end{equation}
Eq. (17) resembles the well-known Friedmann equation in its Newtonian form \citep{Ryden2016} except for the additional radiation pressure term.

\subsection{Distance-redshift relation}
Introducing the Hubble parameter $H(a) = \dot a /a$, eq. (17) can be rewritten in the form
\begin{equation}\label{eq19}
H^2\left(a\right) = H^2_0 \left[{\Omega_m a^{-3} + \Omega_a a^{-5} + \Omega_k a^{-2}}\right] \,.
\end{equation}
where
\begin{equation}\label{eq20}
H^2_0 = \frac{8}{3} \pi G \rho_0 - \frac{2}{3 \pi} \frac{j_0}{c \rho_0 R_0} + 2 U \,, 
\end{equation}
is the Hubble parameter at the present time, and $\Omega_m$, $\Omega_a$ and $\Omega_k$ denote the normalized gravity, radiation-absorption and energy terms, 
\begin{equation}\label{eq21}
\begin{split}
\Omega_m &= \frac{1}{H^2_0} \left({ \frac{8}{3} \pi G \rho_0}\right) ,\\ 
\Omega_a &= -\frac{1}{H^2_0} \left({\frac{2}{3 \pi} \frac{j_0}{c \rho_0 R_0}}\right) ,\\
\Omega_k &= \frac{2 U}{H^2_0} , 
\end{split}
\end{equation}
with
\begin{equation}\label{eq22}
\Omega_m + \Omega_a + \Omega_k = 1 \,. 
\end{equation}

The scale factor $a$ of the Universe with the zero expansion rate is defined by 
\begin{equation}\label{eq23}
H^2\left(a\right) = 0 \,, 
\end{equation}
which yields a cubic equation in $a$
\begin{equation}\label{eq24}
\Omega_k a^3 + \Omega_m a^2 + \Omega_a = 0 \,. 
\end{equation}
Taking into account that $\Omega_m > 0$ and $\Omega_a < 0$, eq. (24) has two distinct real positive roots for 
\begin{equation}\label{eq25}
\left({\frac{\Omega_m}{3}}\right)^2 > \left({\frac{\Omega_k}{2}}\right)^2 \left|\Omega_a\right| \,, 
\end{equation}
and
\begin{equation}\label{eq26}
\Omega_k < 0 \,. 
\end{equation}

Negative $\Omega_a$ and $\Omega_k$ imply that
\begin{equation}\label{eq27}
\Omega_m > 1 \,. 
\end{equation}
and
\begin{equation}\label{eq28}
\rho_0 > \rho_c = \frac{8 \pi G}{3 H^2_0} \,. 
\end{equation}

Under these conditions, eq. (19) describes a universe with a cyclic expansion/contraction history and the two real positive roots $a_\mathrm{min}$ and $a_\mathrm{max}$ define the minimum and maximum scale factors of the Universe. For $\Omega_a \ll 1$, the scale factors $a_\mathrm{min}$ and $a_\mathrm{max}$ read approximately
\begin{equation}\label{eq29}
a_\mathrm{min} \cong \sqrt{\left|{\frac{\Omega_a}{\Omega_m}}\right|} \,\,\, \mathrm{and} \,\,\, 
a_\mathrm{max} \cong       \left|{\frac{\Omega_m}{\Omega_k}}\right| \,, 
\end{equation}
and the maximum redshift is
\begin{equation}\label{eq30}
z_\mathrm{max} = \frac{1}{a_\mathrm{min}} - 1 \,. 
\end{equation}

The scale factors $a$ of the Universe with the maximum expansion/contraction rates are defined by
\begin{equation}\label{eq31}
\frac{d}{da} H^2 \left(a\right) = 0 \,, 
\end{equation}
which yields a cubic equation in $a$
\begin{equation}\label{eq32}
2 \Omega_k a^3 + 3 \Omega_m a^2 + 5 \Omega_a = 0 \,. 
\end{equation}

Taking into account eqs (7) and (21), the deceleration of the expansion reads
\begin{equation}\label{eq33}
\ddot a = -\frac{1}{2} H^2_0 \left[\Omega_m a^{-2} + 3 \Omega_a a^{-4}\right] \,. 
\end{equation}
Hence, the zero deceleration is for the scale factor
\begin{equation}\label{eq34}
a = \sqrt{\left|{\frac{3\Omega_a}{\Omega_m}}\right|} \,. 
\end{equation}

Finally, the comoving distance as a function of redshift is expressed from eq. (19) as follows
\begin{equation}\label{eq35}
\begin{split}
dr = \frac{c}{H_0} \frac{dz}{\sqrt{\Omega_m \left(1+z\right)^3 +  
\Omega_a \left(1+z\right)^5 + \Omega_k \left(1+z\right)^2}} \,. 
\end{split}
\end{equation}
%

\section{Parameters for modeling}

For calculating the expansion history and cosmic dynamics of the Universe, we need observations of cosmological parameters such as the distribution of matter in galaxies and IGM, the galaxy luminosity density, the intensity of the EBL, the mean galactic and intergalactic opacities, and the expansion rate of the Universe at the present time.

\subsection{EBL and the galaxy luminosity density}
The EBL covers a wide range of wavelengths from 0.1 to 1000 μm. It was measured, for example, by the IRAS, FIRAS, DIRBE on COBE, and SCUBA instruments (for reviews, see \citet{Hauser2001, Lagache2005, Cooray2016}. The direct measurements are supplemented by integrating light from extragalactic source counts \citep{Madau2000, Hauser2001} and by attenuation of gamma rays from distant blazars due to scattering on the EBL \citep{Kneiske2004, Dwek2005, Primack2011, Gilmore2012}. The EBL spectrum has two maxima: associated with the radiation of stars (at $0.7 - 2 \mu$m) and with the thermal radiation of dust in galaxies (at $100 - 200 \mu$m), see \citet{Schlegel1998, Calzetti2000}. Despite extensive measurements, uncertainties of the EBL are still large. The total EBL should fall between 40 and 200 $\mathrm{n W\, m^{-2}\,sr^{-1}}$ \citep[his fig. 1]{Vavrycuk2018} with the most likely value \citep{Hauser2001, Bernstein2002a, Bernstein2002b, Bernstein2002c, Bernstein2007}
\begin{equation}\label{eq36}
I^\mathrm{EBL} = 80 - 100 \,\, \mathrm{nW\,m^{-2}\,sr^{-1}} \,.
\end{equation}

The galaxy luminosity density is determined from the Schechter function \citep{Schechter1976}. It has been measured by large surveys 2dFGRS \citep{Cross2001}, SDSS \citep{Blanton2001, Blanton2003} or CS \citep{Geller1997, Brown2001}. The luminosity function in the R-band was estimated at $z = 0$ to be $\left(1.84 \pm 0.04\right) \times 10^8 \,\,h \, L_\odot \,\, \mathrm{Mpc^{-3}}$ for the SDSS data \citep{Blanton2003} and $\left(1.9 \pm 0.6\right) \times 10^8 \,\, h \, L_\odot \,\, \mathrm{Mpc^{-3}}$ for the CS data \citep{Brown2001}. The bolometric luminosity density is estimated by considering the spectral energy distribution (SED) of galaxies averaged over different galaxy types, being thus $1.4 - 2.0$ times larger than that in the R-band \citep[his table 2]{Vavrycuk2017a}
\begin{equation}\label{eq37}
j = 2.5 - 3.8 \times 10^8 \,\, h\, L_\odot \, \mathrm{Mpc^{-3}} \,.
\end{equation}
%

\subsection{Galactic and intergalactic opacity}

The opacity of galaxies due to extinction of light by dust is strongly variable (for a review, see \citet{Calzetti2001}). The most transparent galaxies are ellipticals with an effective extinction $A_V$ of $0.04 - 0.08$ mag. The opacity of spiral and irregular galaxies is higher. The opacity of the disk in the face-on view is formed by two components \citep{Holwerda2005a}: an optically thicker component ($A_I = 0.5 - 4$ mag) associated with the spiral arms and a relatively constant optically thinner disk ($A_I = 0.5$ mag). The inclination-averaged extinction is typically: $0.5 - 0.75$ mag for Sa-Sab galaxies, $0.65 - 0.95$ mag for the Sb-Scd galaxies, and $0.3 - 0.4$ mag for the irregular galaxies at the B-band \citep{Calzetti2001}. Taking into account a relative distribution of galaxy types in the Universe \citep[his table 2]{Vavrycuk2016}, the overall mean galactic visual opacity $\kappa_V$ is about $0.22 \pm 0.08$ \citep{Vavrycuk2017a}.

The intergalactic opacity is strongly spatially dependent being mostly appreciable at close distance from galaxies and in intracluster space. Measurements of \citet{Menard2010a} indicate that the visual intergalactic attenuation is $A_V = (1.3 \pm 0.1) \times 10^{-2}$ mag at a distance from a galaxy of up to 170 kpc and $A_V = (1.3 \pm 0.3) \times 10^{-3}$ mag at a distance of up to 1.7 Mpc. Similar values are observed for the visual attenuation of intracluster dust \citep{Muller2008, Chelouche2007}. An averaged value of intergalactic extinction was measured by \citet{Menard2010a} by correlating the brightness of $\sim 85.000$ quasars at $z > 1$ with the position of 24 million galaxies at $z \sim 0.3$ derived from the Sloan Digital Sky Survey. The authors estimated $A_V$ to about 0.03 mag at $z = 0.5$. A consistent opacity was reported by \citet{Xie2015} who studied the luminosity and redshifts of the quasar continuum of $\sim 90.000$ objects and estimated the intergalactic opacity at $z < 1.5$ as 
\begin{equation}\label{eq38}
A_V \sim 0.02 \,h \, \mathrm{Gpc}^{-1} \,.
\end{equation}

Dust extinction can also be measured from the hydrogen column densities of damped Lyman $\alpha$ absorbers (DLAs). Based on the Copernicus data, \citet{Bohlin1978} reports a linear relationship between the total hydrogen column density, $N_\mathrm{H} = 2 N_\mathrm{H2} + N_\mathrm{HI}$, and the color excess and visual attenuation 
\begin{equation}\label{eq39}
N_\mathrm{H} / \left(A_B-A_V\right) = 5.8 \times 10^{21} \, \mathrm{cm}^{-2} \, \mathrm{mag}^{-1} \,, 
\end{equation}
\begin{equation}\label{eq40}
N_\mathrm{H} / A_V \approx 1.87 \times 10^{21} \, \mathrm{cm}^{-2} \, \mathrm{mag}^{-1}\, \, \mathrm{for} \,\, R_V = 3.1\,. 
\end{equation}

The result has been confirmed by \citet{Rachford2002} using the FUSE data who refined the slope in eq. (39) to $5.6 \times 10^{21} \, \mathrm{cm}^{-2} \, \mathrm{mag}^{-1}$. Taking into account observations of the mean cross-section density of DLAs \citep{Zwaan2005}, $\langle n \sigma \rangle =  \left(1.13 \pm 0.15 \right) \times 10^{-5} \, h \, \mathrm{Mpc}^{-1}$, the characteristic column density of DLAs, $N_\mathrm{HI} \sim 10^{21} \, \mathrm{cm}^{-2}$ reported by \citet{Zwaan2005}, and the mean molecular hydrogen fraction in DLAs of about $0.4 - 0.6$  \citep[their Table 8]{Rachford2002}, the intergalactic attenuation $A_V$ at $z = 0$ is from eq. (40):  $A_V \sim 1-2 \times 10^{-5} \, h \, \mathrm{Mpc}^{-1}$, which is the result of \citet{Xie2015}: $A_V \sim 2 \times 10^{-5} \, h \, \mathrm{Mpc}^{-1}$. A low value of intergalactic attenuation indicates that the intergalactic space is almost transparent at $z = 0$. The intergalactic attenuation is, however, redshift dependent. It increases with redshift, and a transparent universe becomes significantly opaque at redshifts of $z > 3$, see \citet{Vavrycuk2017a}.

\subsection{Size and number density of galaxies}

The distribution of galaxies is fairly variable because of galaxy clustering and presence of voids in the universe \citep{Peebles2001, Jones2004, vonBenda_Beckmann2008}. The galaxy number density might be 10 times higher or more in clusters, at distances up to $15 - 20$ Mpc than the density averaged over larger distances. The number density averaged over hundreds of Mpc is, however, stable. It is derived from the Schechter luminosity function \citep{Schechter1976} and lies in the range of $0.010-0.025 \,\, h^3 \,\, \mathrm{Mpc^{-3}}$, reported by \citet{Peebles1993, Peacock1999, Blanton2001, Blanton2003}. These values correspond to the mean distance $R_0$ between galaxies $R_0 = 2.1 - 2.9 \,\,h^{-1} \,\, \mathrm{Mpc}$ calculated using eq. (18).

The size of galaxies depends on their stellar mass, luminosity, and the morphological type. The galaxy luminosities are between $\sim 10^8 \,\, L_\odot$ and $\sim 10^{12}\,\, L_\odot$ with effective radii between $\sim 0.1\,\, h^{-1}$ kpc and $\sim 10\,\, h^{-1}$ kpc. For late-type galaxies, the characteristic luminosity is -20.5 in the R-band \citep{Shen2003}; the Petrosian half-light radius is $\sim 2.5 - 3.0\,\, h^{-1}$ kpc and the $R_{90}$ radius is $R_{90} = 7.5 - 9 \,\, h^{-1}$ kpc \citep{Graham2005}. This value is close to a commonly assumed mean galaxy radius $R = 10 \,\, h^{-1}$ kpc \citep{Peebles1993, Peacock1999}.

\subsection{Size and number density of damped Lyman $\alpha$ clouds}

Light is not absorbed only by dust in galaxies but also by intergalactic dust in the damped Lyman $\alpha$  clouds (DLA). The DLA are massive clouds with column densities $N_\mathrm{HI} > 2 \times 10^{20} \, \mathrm{cm}^{-2}$. They are self-shielded against ionizing radiation from outside and contain mostly neutral hydrogen and form reservoirs of dust \citep{Wolfe2005, Meiksin2009}. The dust content is usually estimated from the abundance ratio [Cr/Zn] assuming that this stable ratio is changed in dusty environment because Cr is depleted on dust grains but Zn is undepleted \citep{Pettini1994, Pettini1997}. Other dust indicators are depletion of Fe and Si relative to S and Zn \citep{Petitjean2002}. Dust in DLAs is also evidenced by reddening measured by comparing composite spectra of quasars with and without foreground absorption systems and by searching for a systematic change of the continuum slope \citep{Fall_Pei1989, Pei1991, Wild2006}. In addition, \citet{Vladilo2006}, \citet{Noterdaeme2017} and \citet{Ma2017} identified a clear signature of the 2175 \AA  absorption bump in quasar spectra produced by dust in intervening DLAs.

The number density of DLAs is considerably higher than that of galaxies. It can be deduced from measurements of the Lyman $\alpha$ (Ly$\alpha$) absorption lines produced in quasar optical spectra by neutral hydrogen (H I) in intergalactic clouds ionized by ultraviolet radiation at wavelength of 1216 \AA. Taking into account the mean cross-section density of DLAs reported by \citet{Zwaan2005} $\langle n \sigma \rangle =  \left(1.13 \pm 0.15 \right) \times 10^{-5} \, h \, \mathrm{Mpc}^{-1}$ and size $R$ of 5-10 kpc \citep{Cooke_OMeara2015}, we get the number density of $0.2 - 0.05 \,\, h^3 \, \mathrm{Mpc^{-3}}$. Since dust is also partially contained in less massive but more abundant LA systems, the number density $0.2 \,\, h^3 \, \mathrm{Mpc^{-3}}$ seems to be more appropriate for numerical modeling if the other less massive LAs are neglected. Subsequently, the mean distance $R_0$ between the DLAs is estimated by eq. (18) to be about 1 Mpc.

\begin{table*}
%
%
\caption{Maximum redshift and scale factor in the cyclic model of the opaque universe}  

\label{Table:1}      
%
%
\centering                          
\begin{tabular}{c c c c c c c c}  
%
%
\hline\hline                 
%
%
 $Model$ & $j_0$ & $R_0$ & $\Omega_m$ & $\Omega_a$ &  $\Omega_k$  & $a_\mathrm{max}$ & $z_\mathrm{max}$\\
 & $ (10^8 \, h\, L_\odot \, \mathrm{Mpc}^{-3})$ & $(h^{-1}\,\mathrm{Mpc})$ &  &  &  &  &  \\
%
%
\hline                        
%
%
A & 8.0 & 0.5 & 1.2 & $\,\, -4.53 \times 10^{-3}$ & $\,\, -0.195$ &  6.1 & 15.2 \\
B & 3.5 & 1.2 & 1.2 & $\,\, -5.90 \times 10^{-4}$ & $\,\, -0.199$ &  6.0 & 44.0 \\    
C & 4.5 & 0.8 & 1.2 & $\,\, -1.45 \times 10^{-3}$ & $\,\, -0.199$ &  6.0 & 27.7 \\
D & 4.5 & 0.8 & 1.1 & $\,\, -1.45 \times 10^{-3}$ & $\,\, -0.099$ & 11.2 & 26.5 \\
E & 4.5 & 0.8 & 1.3 & $\,\, -1.45 \times 10^{-3}$ & $\,\, -0.299$ &  4.4 & 28.8 \\
%
%
\hline                                  
\end{tabular}
%
%
\footnotetext{
The Hubble constant is $H_0 = 73.3 \,\, \mathrm{km \, s^{-1} \, Mpc^{-1}}$. 
Models A, B and C predict low, high and optimum values of $z_\mathrm{max}$. 
Models E, D and C predict low, high and optimum values of $a_\mathrm{max}$. 
\\}
\end{table*}

\subsection{Mass density of the Universe}

A simplest and most straightforward method how to estimate the mass density is based on galaxy surveys and computation of the mass from the observed galaxy luminosity and from the mass-to-light ratio ($M/L$) that reflects the total amount of the mass relative to the light within a given scale. The $M/L$ ratio is, however, scale dependent and increases from bright, luminous parts of galaxies to their halos (with radius of $\sim 200$ kpc) formed by (baryonic and/or speculative non-baryonic) dark matter. The $M/L$ ratio depends also on a galaxy type being about 3 to 4 times larger for elliptical/SO galaxies than for typical spirals, hence the observed $M/L_B$ is $\sim 100 \, h$ for spirals, but $\sim 400 \, h$ for ellipticals at radius of $\sim 200$ kpc, see \citet{Bahcall1995}. Considering the mean asymptotic ratio $M/L_B$ of $200 - 300 \, h$ and the observed mean luminosity density of the universe at $z = 0$ of $\sim 2.5 \times 10^8 \,\, h \, L_\odot \, \mathrm{Mpc^{-3}}$ reported by \citet{Cross2001}, the mass density $\Omega_m$ associated with galaxies is about $0.2 - 0.3$ ($\Omega_m = 1$ means the critical density).

Another source of matter in the universe is connected to Ly$\alpha$ absorbers containing photoionized hydrogen at $\sim 10^4$ K and being detected by the Ly$\alpha$ forest in quasar spectra \citep{Meiksin2009}. These systems are partly located in the galaxy halos, but a significant portion of them cannot be associated to any galaxy, being observed, for example, in voids \citep{Penton2002, Tejos2012, Tejos2014}. The Ly$\alpha$ absorbers also form the intragroup and intracluster medium \citep{Bielby2017} and the intergalactic medium nearby the other large-scale galaxy structures like the galaxy filaments \citep{Tejos2014, Wakker2015}. In addition, it is speculated that a large amount of matter is located in the warm-hot intergalactic medium (WHIM) that is a gaseous phase of moderate to low density ($\sim 10 - 30$ times the mean density of the Universe) and at temperatures of $10^5 - 10^7$ K. Although it is difficult to observe the WHIM because of low column densities of HI in the hot gas, they might be potentially detected by surveys of broad HI Ly$\alpha$ absorbers (BLAs) as reported by \citet{Nicastro2018} or \citet{Pessa2018}. 

Hence, we can conclude that the estimate of mass density $\Omega_m = 0.2 - 0.3$ inferred from distribution of galaxies is just a lower limit, while the upper limit of $\Omega_m$ is unconstrained being possibly close to or even higher than 1. This statement contradicts the commonly accepted value of $\Omega_m = 0.3$ reported by \citet{Ade2014b, Planck_2015_XIII, Planck_2018_VI} which is based on the interpretation of the CMB as a relic radiation of the Big Bang. Since the CMB is interpreted as thermal radiation of intergalactic dust \citep{Vavrycuk2017a, Vavrycuk2018}, the results of \citet{Ade2014b, Planck_2015_XIII, Planck_2018_VI} are irrelevant for the presented theory and not considered in the numerical modeling in this paper.

\subsection{Hubble constant $H_0$}

Among numerous approaches used for measuring the present-day expansion of the Universe I focus on direct methods, which are considered to be most reliable (for a review, see \citet{Jackson2015}). These methods are based on measuring local distances up to $20 - 30$ Mpc using Cepheid variables observed by the Hubble Space Telescope (HST). The HST galaxies with distance measured with the Cepheid variables are then used to calibrate type Ia supernovae (SNe). With this calibration, the distance measure can be extended to other more distant galaxies (hundreds of Mpc) in which SNe Ia are detected \citep{Freedman2001, Riess2011}. The best estimate of $H_0$ obtained by \citet{Riess2016} is $73.25 \pm 1.74 \,\, \mathrm{km \, s^{-1} \, Mpc^{-1}}$. The precision of the distance scale was further reduced by a factor of 2.5 by \citet{Riess2018}. However, this value and its accuracy should be taken with reservations, because it comes from measurements uncorrected for intergalactic opacity, see Section 5 Comparison with supernovae measurements. The other methods based, e.g., on the Sunyaev-Zel'dovich effect \citep{Birkinshaw1999, Carlstrom2002, Bonamente2006} or gravitational lensing \citep{Suyu2013, Bonvin2017} yield consistent values of $H_0$ but with a lower accuracy. The exception is $H_0$ determined from interpretation of acoustic peaks in the CMB spectrum provided by \citet{Aghanim2016}, which is claimed to have a high accuracy with a value of $H_0 = 67.3 \pm 1.2 \,\, \mathrm{km \, s^{-1} \, Mpc^{-1}}$. The results based on the interpretation of the CMB as a relic radiation of the Big Bang are not, however, considered in this paper.

\section{Results}

\begin{figure}
\includegraphics[angle=0,width=7.0cm,trim=20 10 60 10]{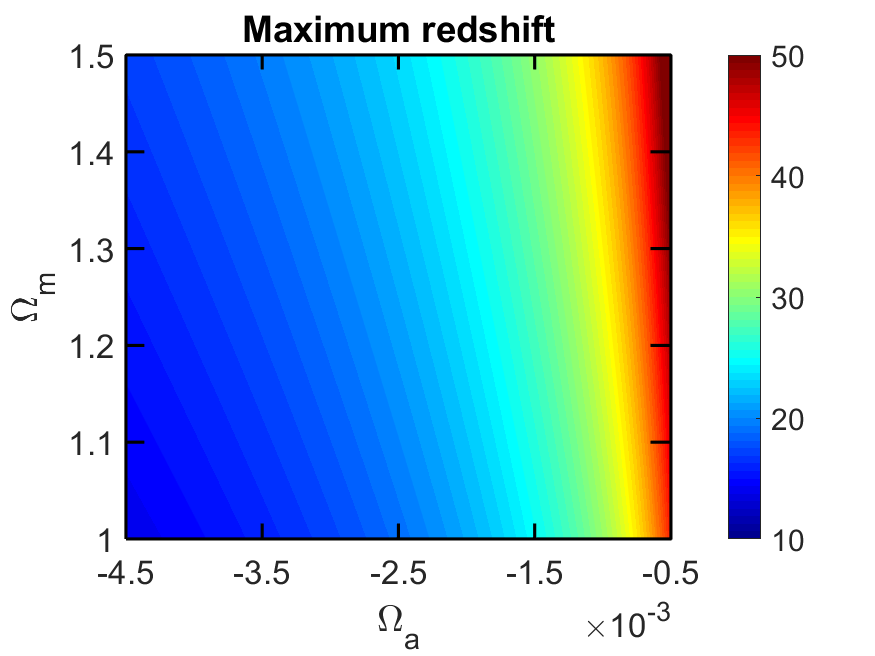}
\caption{
Maximum redshift as a function of $\Omega_m$ and $\Omega_a$. 
}
\label{fig:2}
\end{figure}

\begin{figure}
\includegraphics[angle=0,width=7.5cm,trim=100 40 160 100]{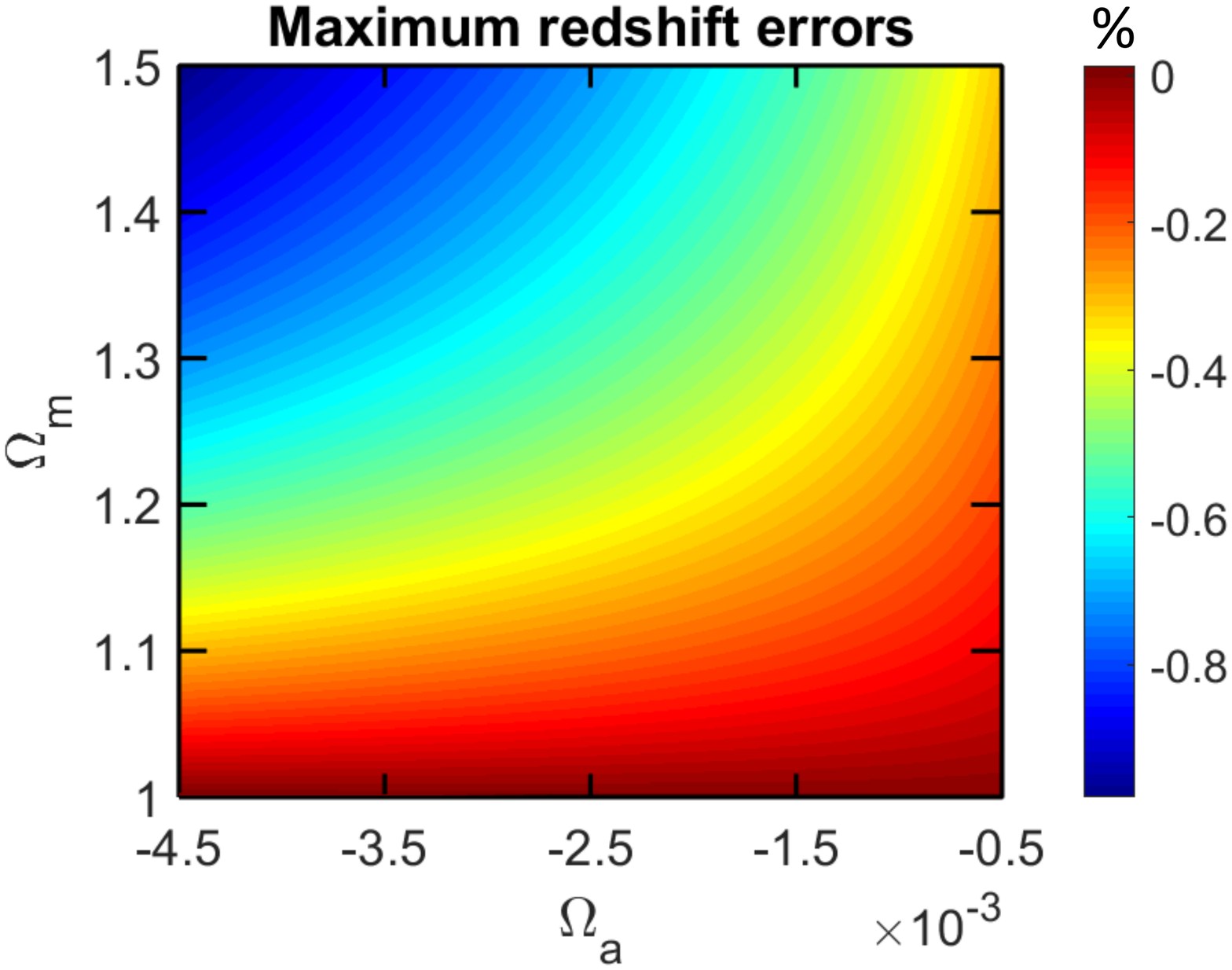}
\caption{
Errors of the approximate maximum redshift as a function of $\Omega_m$ and $\Omega_a$. The approximate values are calculated by eqs (29) and (30). The errors are evaluated in \%.
}
\label{fig:3}
\end{figure}

\begin{figure*}
\includegraphics[angle=0, width=13cm, trim=60 180 180 150]{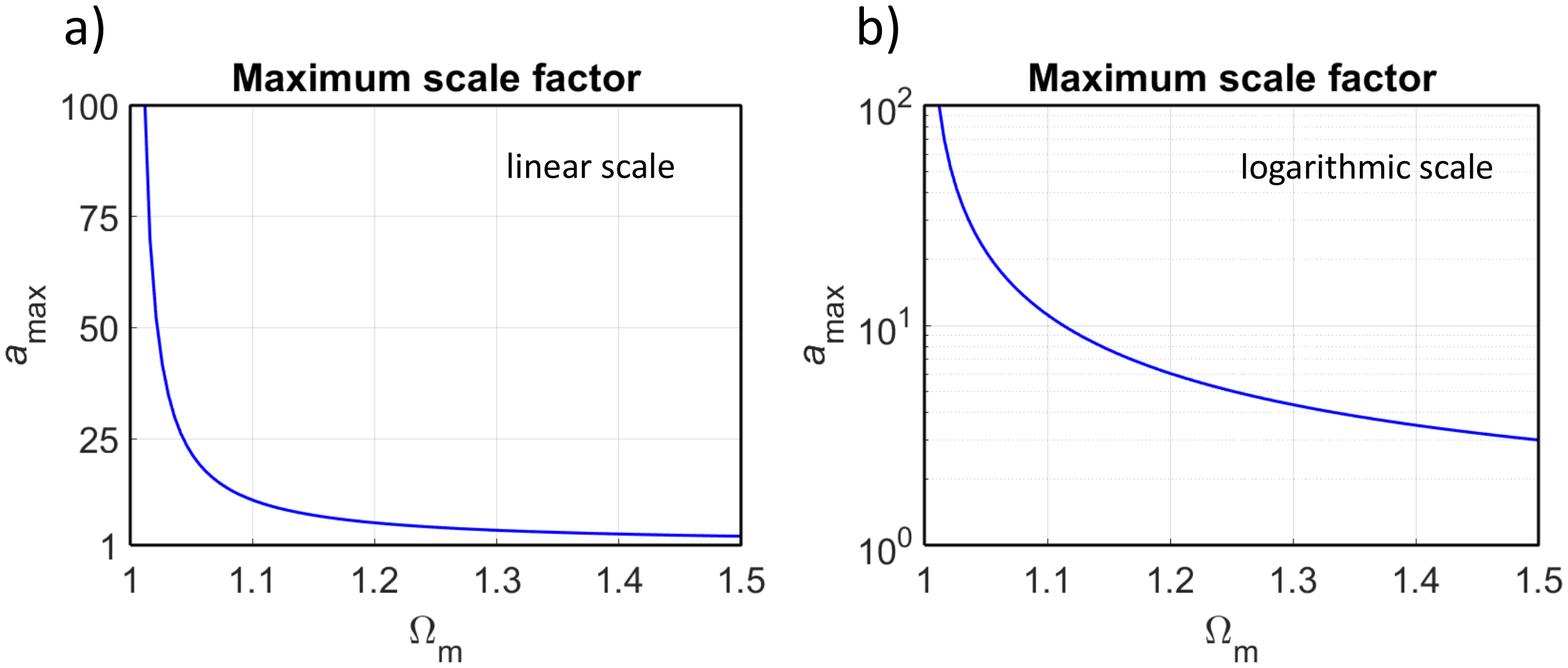}
\caption{
The maximum scale factor as a function of $\Omega_m$. (a) Linear scale, (b) logarithmic scale. The dependence on $\Omega_a$ is negligible (for values considered in Fig.~\ref{fig:2}).
}
\label{fig:4}
\end{figure*}

\begin{figure*}
\includegraphics[angle=0,width=13 cm, trim=60 120 180 100]{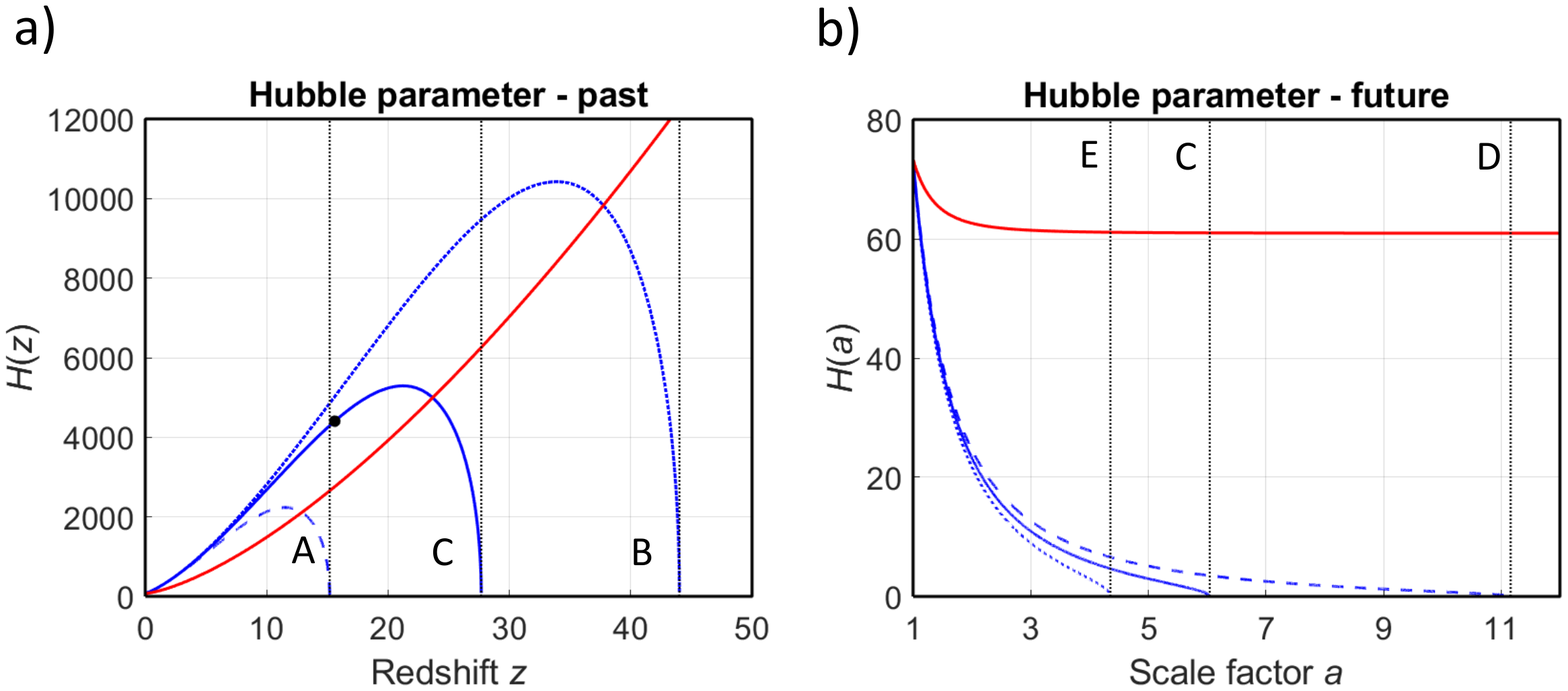}
\caption{
The evolution of the Hubble parameter with redshift in the past and with the scale factor in future (in $\mathrm{km \, s^{-1} \, Mpc^{-1}}$). (a) The blue dashed, dotted and solid lines show Models A, B and C in Tab. 1. (b) The blue solid, dashed, and dotted lines show Models C, D and E in Tab. 1. The black dotted lines mark the predicted maximum redshifts (a) and maximum scale factors (b) for the models considered. The black dot denotes the state in C when the deceleration of the expansion is zero. The dot is not at the maximum of $H(z)$ because the zero deceleration is with respect to time but not with respect to $z$. The red solid line shows the flat $\Lambda$CDM model with $\Omega_m = 0.3$, $\Omega_\Lambda = 0.7$. The Hubble constant $H_0$ is $73.3 \,\, \mathrm{km \, s^{-1} \, Mpc^{-1}}$.
}
\label{fig:5}
\end{figure*}

\begin{figure}
\includegraphics[angle=0,width=7.0cm,trim=40 10 60 10]{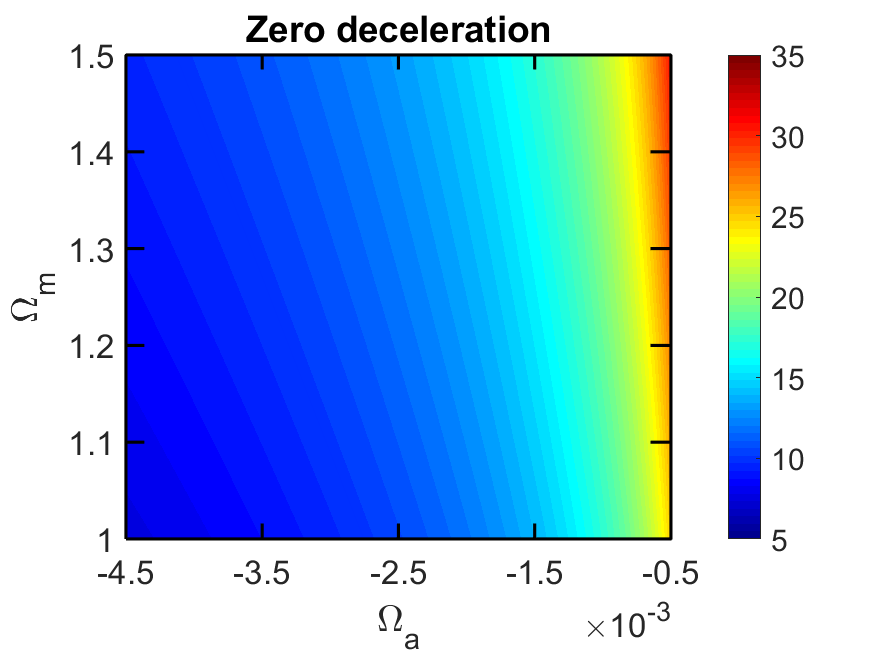}
\caption{
Redshift $z$ for the zero deceleration as a function of $\Omega_m$ and $\Omega_a$. 
}
\label{fig:6}
\end{figure}

\begin{figure}
\includegraphics[angle=0,width=6.5cm,trim= 120 10 180 100]{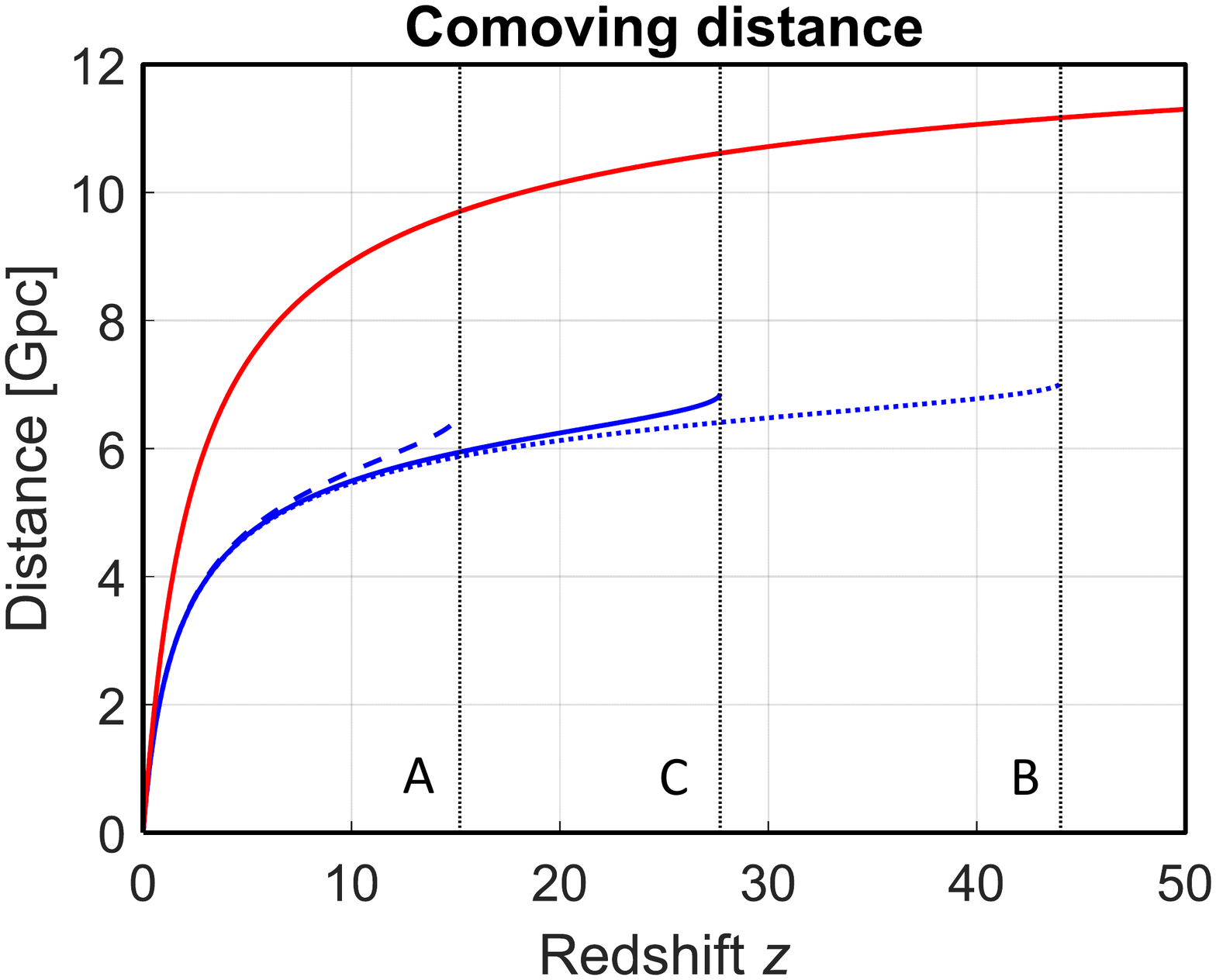}
\caption{
Comoving distance as a function of redshift $z$. The blue dashed, dotted and solid lines show Models A, B and C in Tab. 1. The black dotted lines mark the predicted maximum redshifts for the models considered. The red solid line shows the flat $\Lambda$CDM model with $\Omega_m = 0.3$, $\Omega_\Lambda = 0.7$. The Hubble constant $H_0$ is $73.3 \,\, \mathrm{km \, s^{-1} \, Mpc^{-1}}$.
}
\label{fig:7}
\end{figure}

Estimating cosmological parameters and their uncertainties from observations (see Table 1), we can calculate the upper and lower limits of the volume of the Universe and the evolution of the Hubble parameter with time. Since we confine ourselves to cyclic cosmological models only, the mass density of the Universe higher than the critical density is considered, and subsequently $\Omega_m$ is higher than 1. Parameter $\Omega_a$ varies from $-4.5 \times 10^{-3}$ to $-5.9 \times 10^{-4}$ depending on the luminosity density $j_0$ and mean distance $R_0$ between DLAs (see eq. (21) and Table 1). 

As seen from Fig.~\ref{fig:2}, the maximum redshift of the Universe depends mostly on $\Omega_a$, and ranges from 15 to 45. The maximum redshift $z_\mathrm{max}$ corresponding to a minimum scale factor $a_\mathrm{min}$ calculated by approximate eq. (29) has an accuracy higher than 1\% (see Fig.~\ref{fig:3}). In contrast to $a_\mathrm{min}$ depending mostly on $\Omega_a$, the maximum scale factor $a_\mathrm{max}$ of the Universe depends primarily on $\Omega_m$. Fig.~\ref{fig:4} shows that $a_\mathrm{max}$ rapidly decreases with increasing $\Omega_m$. Obviously, the limiting value is $\Omega_m = 1$, when $a_\mathrm{max}$ is infinite. For $\Omega_m = 1.1$, 1.2, 13 and 1.5, the scale factor $a_\mathrm{max}$ is 11.1, 6.0, 4.4 and 3.0, respectively.

The history of the Hubble parameter $H(z)$ and its evolution in future $H(a)$ calculated by eq. (19) is shown in Fig.~\ref{fig:5} for five scenarios summarized in Table 1. As mentioned, the form of $H(z)$ is controlled by $\Omega_a$ (Fig.~\ref{fig:5}a), while the form of $H(a)$ is controlled by $\Omega_m$ (Fig.~\ref{fig:5}a). The Hubble parameter $H(z)$ increases with redshift up to its maximum. After that the function rapidly decreases to zero. The drop of $H(z)$ is due to a fast increase of light attenuation producing strong repulsive forces at high redshift. For future epochs, function $H(a)$ is predicted to monotonously decrease to zero. The rate of decrease is controlled just by gravitational forces and the energy of the system. The repulsive forces originating in light attenuation are negligible. For a comparison, Fig.~\ref{fig:5} also shows the Hubble parameter $H(a)$ for the $\Lambda$CDM model \citep{Planck_2015_XIII} which is the standard model of the Big Bang cosmology characterized by the accelerating expansion
\begin{equation}\label{eq41}
H^2\left(a\right) = H^2_0 \left[{\Omega_m a^{-3} + \Omega_\Lambda + \Omega_k a^{-2}}\right] \,.
\end{equation}
with $\Omega_m = 0.3$, $\Omega_\Lambda = 0.7$ and $\Omega_k = 0$.

As regards the deceleration of the expansion, it becomes zero before $H(z)$ attaining its maximum (see the black dot in Fig.~\ref{fig:5}). The redshift of the zero deceleration is about 3/4 of the maximum redshift of the Universe. As seen from Fig.~\ref{fig:6}, the redshift of the zero deceleration is less than 15 for most of considered values of $\Omega_m$ and $\Omega_a$.

Also the distance-redshift relation is quite different for the standard $\Lambda$CDM model and for the proposed cyclic model of the Universe (see Fig.~\ref{fig:7}). In both models, the comoving distance monotonously increases with redshift. However, the redshift can go possibly to infinity in the standard cosmology, while the maximum redshift in the cyclic cosmology is finite. Also the increase of distance with redshift is remarkably steeper for the standard model than for the cyclic models. The ratio between distances in the cyclic and standard cosmological models is about 0.6.

\section{Comparison with supernovae measurements}

\begin{figure*}
\includegraphics[angle=0,width=11 cm,trim=10 40 200 120]{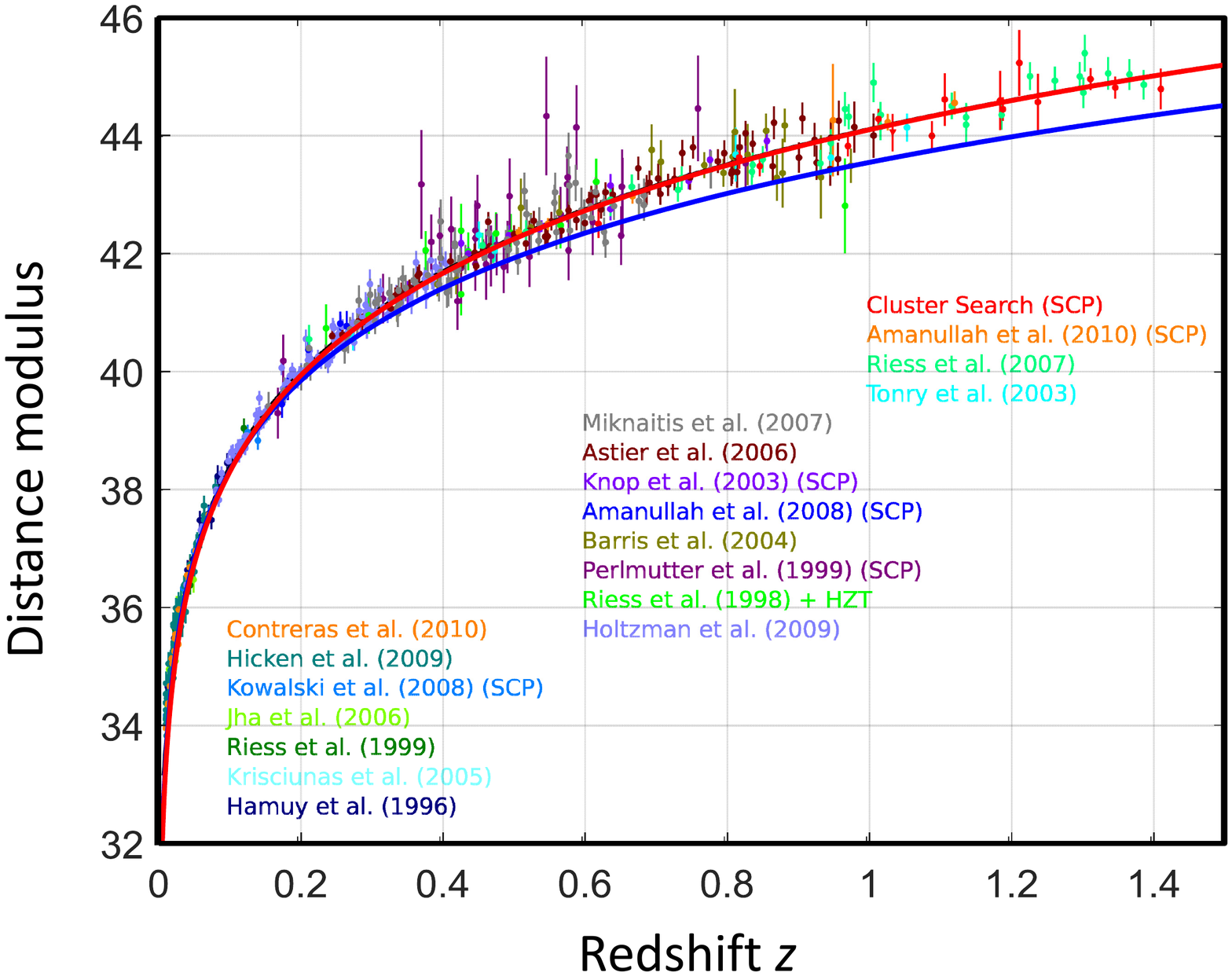}
\caption{
The Hubble diagram with SNe Ia measurements (Union2.1 dataset). Solid red line - the flat $\Lambda$CDM model with $\Omega_m = 0.3$, $\Omega_\Lambda = 0.7$, $\Omega_k = 0$. Solid blue line - the cosmological model with $\Omega_m = 1.2$, $\Omega_\Lambda = 0$, $\Omega_k = -0.2$. The data are not corrected for intergalactic opacity. Modified after \citet[their fig. 4]{Suzuki2012}.
}
\label{fig:8}
\end{figure*}

Measuring the SNe Ia is currently considered as the most accurate method for determining cosmic distances. The uniformity of the SNe Ia allows to use them as the standard candle over a wide distance range and to map directly the expansion history. The first results published by \citet{Riess1998} and \citet{Perlmutter1999} were based on measurements of 16 and 42 high-redshift SNe Ia, respectively. They revealed an unexpected dimming of SNe, which was interpreted by an accelerating expansion. This surprising discovery motivated large-scale systematic searches for SNe Ia and resulted in a rapid extension of supernovae compilations (Fig.~\ref{fig:8}), which comprise now of about one thousand SNe Ia discovered and spectroscopically confirmed \citep{Sullivan2011, Suzuki2012, Campbell2013, Betoule2014, Rest2014, Riess2018}. 

\begin{figure}
\includegraphics[angle=0,width=7.5cm, trim = 130 40 150 150]{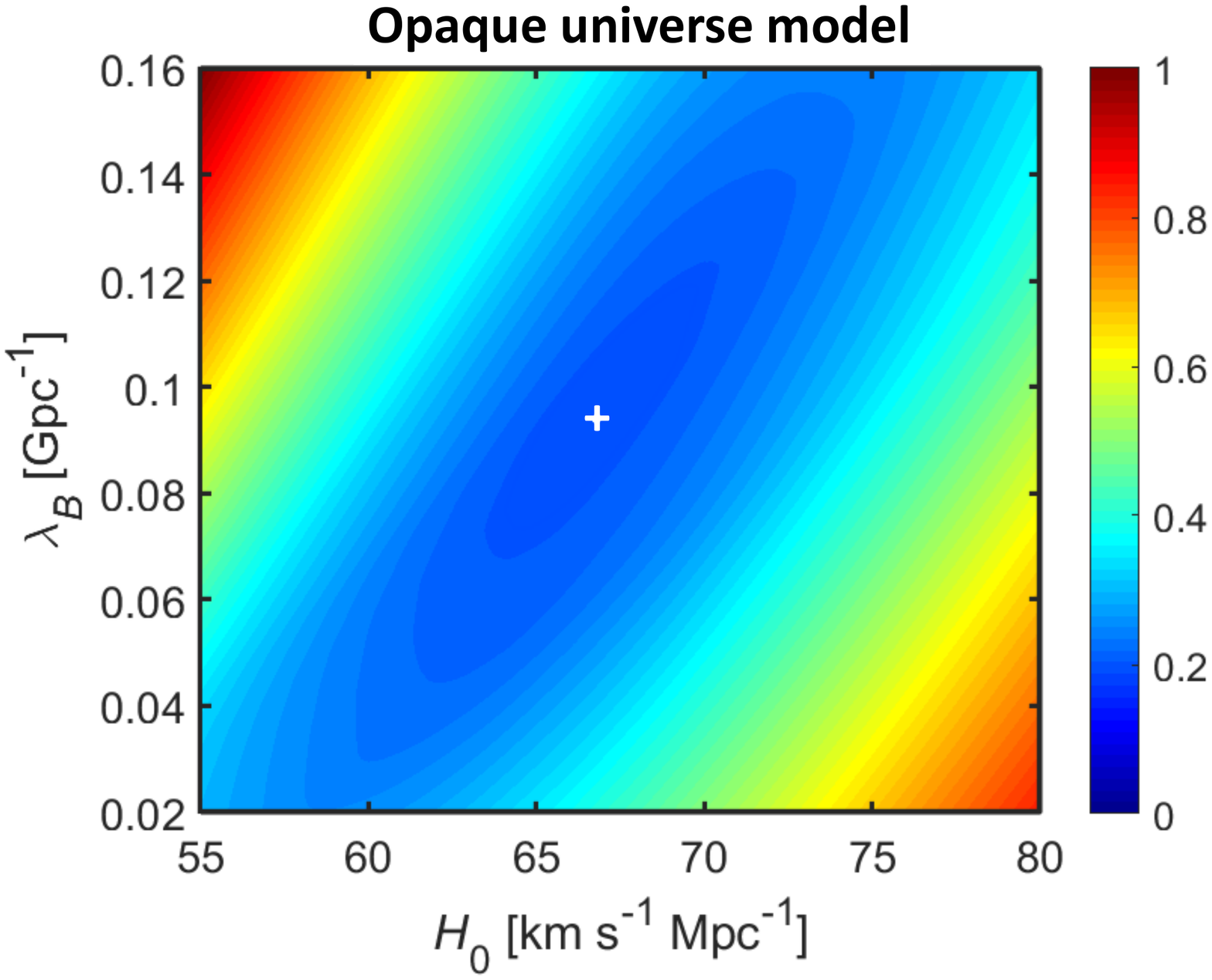}
\caption{
Inversion for optimum cosmological parameters from SNe measurements: opaque universe model. The sum of the distance modulus residua between the predicted model and measurements of the SNe Union2.1 compilation is shown as a function of the B-band intergalactic opacity $\lambda_B$ and the Hubble constant $H_0$. In order the residua to have the same weight for different redshift intervals with a different number of SNe measurements, the sum of absolute values of the residua are calculated in bins ($0 < z < 1.4$) with step of 0.2. Only the most accurate measurements with an error less than 0.25 mag are used. The optimum solution marked by the plus sign is defined by $H_0 = 66.8 \,\, \mathrm{km \, s^{-1} \, Mpc^{-1}}$ and $\lambda_B = 0.094 \,\, \mathrm{Gpc^{-1}}$. 
}
\label{fig:9}
\end{figure}

\begin{figure}
\includegraphics[angle=0,width=7.5cm, trim = 130 0 150 150]{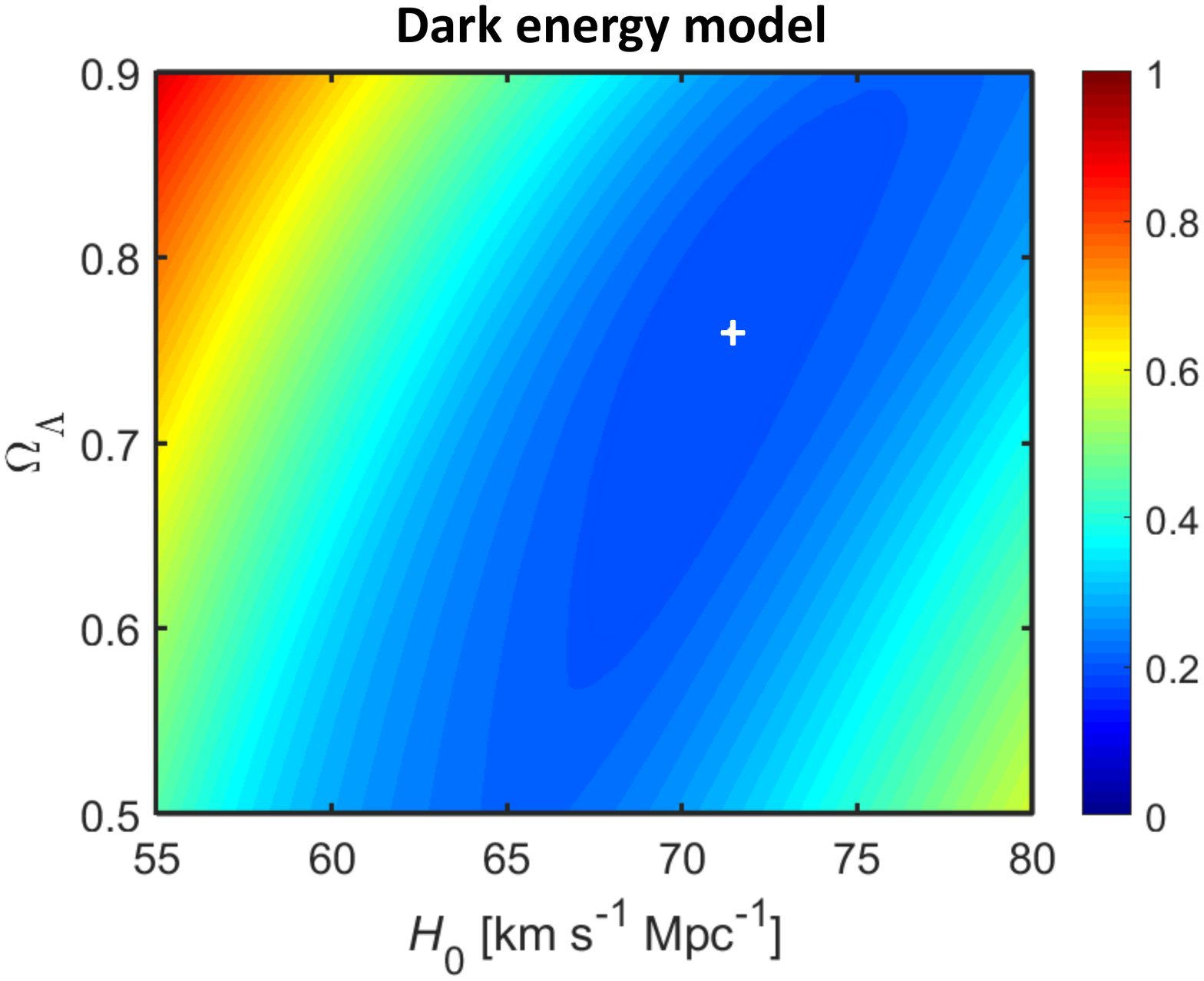}
\caption{
Inversion for optimum cosmological parameters from SNe measurements: flat $\Lambda$CDM model. The sum of the distance modulus residua between the predicted model and measurements of the SNe Union2.1 compilation is shown as a function of $\Omega_\Lambda$ and the Hubble constant $H_0$. In order the residua to have the same weight for different redshift intervals with a different number of SNe measurements, the sum of absolute values of the residua are calculated in bins ($0 < z < 1.4$) with step of 0.2. Only the most accurate measurements with an error less than 0.25 mag are used. The optimum solution marked by the plus sign is defined by $H_0 = 71.4 \,\, \mathrm{km \, s^{-1} \, Mpc^{-1}}$ and $\Omega_\Lambda = 0.76$. 
}
\label{fig:10}
\end{figure}

\begin{figure*}
\includegraphics[angle=0,width=13cm,trim=60 180 200 100]{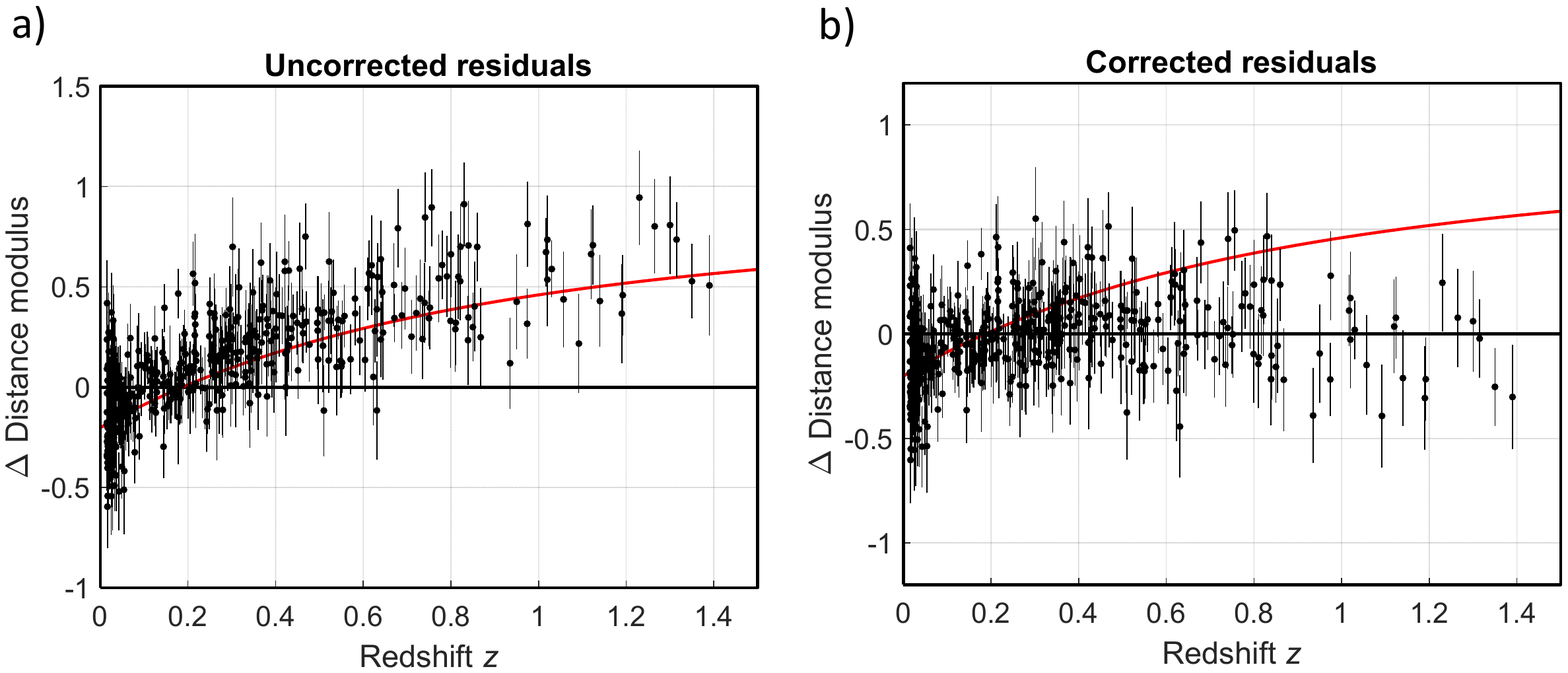}
\caption{
Residual Hubble plots for the SNe Union2.1 compilation. (a) Residuals uncorrected for intergalactic opacity $\lambda_B$. (b) Residuals corrected for intergalactic opacity. The zero line shows the opaque universe model with $\lambda_B = 0.094 \,\, \mathrm{Gpc^{-1}}$, $\Omega_m = 1.2$ and $H_0 = 66.8 \,\, \mathrm{km \, s^{-1} \, Mpc^{-1}}$. The red line shows the flat $\Lambda$CDM model with $H_0 = 71.4 \,\, \mathrm{km \, s^{-1} \, Mpc^{-1}}$, $\Omega_\Lambda = 0.76$ and $\Omega_m = 0.24$. Only the most accurate measurements with an error less than 0.25 mag are shown. Data are taken from \citet{Suzuki2012}.
}
\label{fig:11}
\end{figure*}

\begin{figure*}
\includegraphics[angle=0,width=13cm,trim=60 180 180 60]{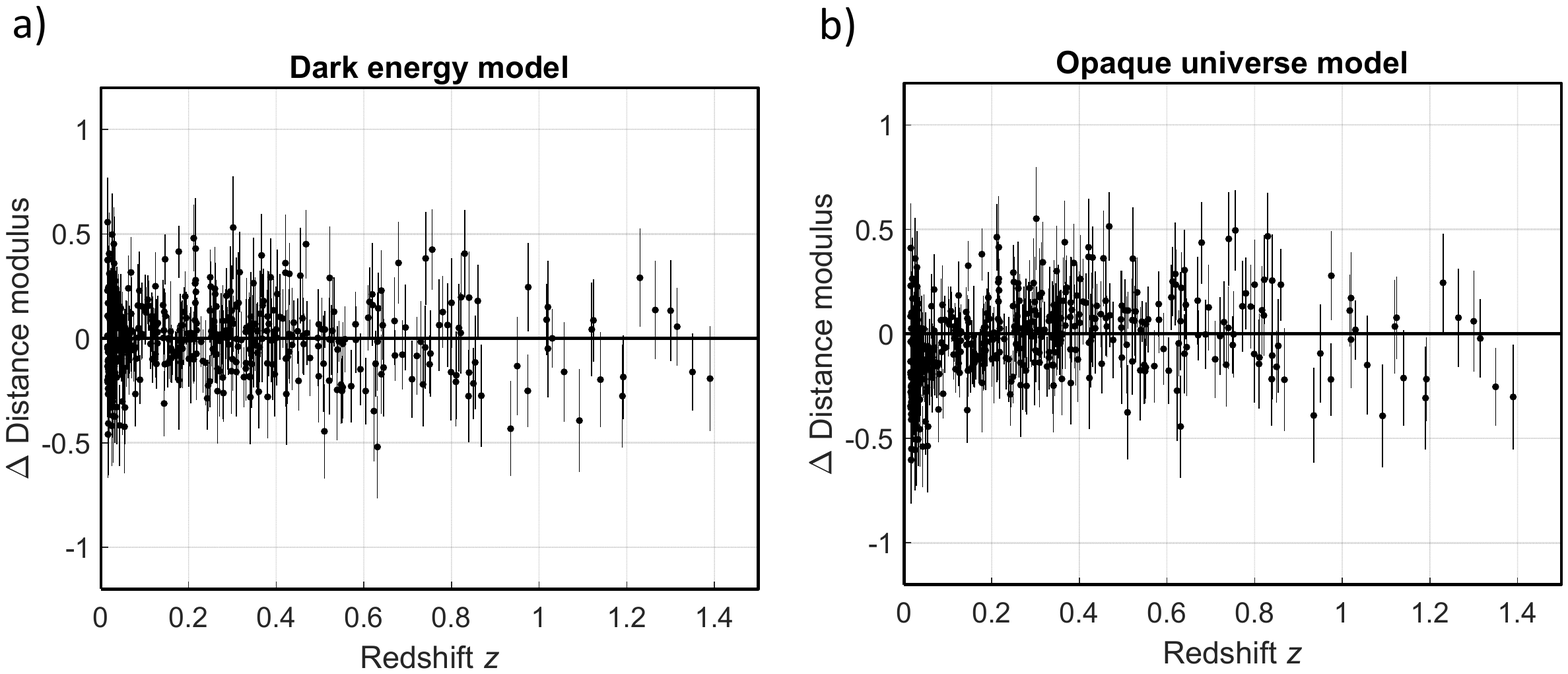}
\caption{
Residual Hubble plots for the SNe Union2.1 compilation. (a) Residuals uncorrected for intergalactic opacity $\lambda_B$. The zero line shows the flat $\Lambda$CDM model with $H_0 = 71.4 \,\, \mathrm{km \, s^{-1} \, Mpc^{-1}}$, $\Omega_\Lambda = 0.76$ and $\Omega_m = 0.24$. (b) Residuals corrected for intergalactic opacity $\lambda_B$. The zero line shows the opaque universe model (model C in Table 1) with $\lambda_B = 0.094 \,\, \mathrm{Gpc^{-1}}$, $\Omega_m = 1.2$, $ \Omega_k = -0.199$ and $H_0 = 66.8  \,\, \mathrm{km \, s^{-1} \, Mpc^{-1}}$. Only the most accurate measurements with an error less than 0.25 mag are shown. Data are taken from \citet{Suzuki2012}.
}
\label{fig:12}
\end{figure*}

Prior to interpretations of the SNe data, corrections must be applied to transform an observed-frame magnitude to a rest-frame magnitude. This includes a cross-filter $K$-correction, a correction for light extinction due to absorption by dust in the host galaxy and in our Galaxy \citep{Perlmutter1999, Nugent2002, Riess2004b}. The uncertainties in the extinction corrections are in general much higher than those in the $K$-corrections \citep{Nugent2002}. Moreover, the uncertainties increase due to neglecting extinction by intergalactic dust which might display different reddening than the interstellar dust. Obviously, these uncertainties question whether the observations of supernovae dimming are not partly or fully a product of intergalactic extinction. 

Intergalactic opacity as a possible origin of dimming of the SNe luminosity was proposed by \citet{Aguirre1999a, Aguirre1999b}. Also \citet{Menard2010b} point out that a reddening-based correction is not sensitive to intergalactic opacity and it might bias the calculated distance modulus and the cosmological parameters describing the accelerating expansion. This problem was also addressed by \citet{Riess2004b} who admitted that some models of intergalactic dust produced a similar dimming as that interpreted by the accelerating expansion. Moreover, recent detailed studies of extinction in SNe spectra revealed that standard extinction corrections are too simplified, because unexpected complexities were detected in reddening, being characterized by a variety of extinction laws with $R_V$ going down to $\sim 1.4$, see \citet{Amanullah2014, Amanullah2015, Gao2015}. The origin of extremely low $R_V$ values is not clear, but it is speculated that the low $R_V$ might indicate small dust grains. Alternatively, multiple scattering on standard dust grains in the circumstellar medium surrounding SNe can cause the effective extinction law to steepen \citep{Goobar2008, Bulla2018}. 

Since the current measurements of supernovae distances ignore the systematic differences between the circumstellar, interstellar and intergalactic dust absorption and correct uniformly for the Galaxy-type extinction (the CCM-law from \citet{Cardelli1989} with $R_V = 3.1$), any systematic trend in the distance-redshift function caused by intergalactic light extinction is suppressed. This might be the reason why any firm evidence for a residual redshift-dependent evolution has not been found yet \citep{Betoule2014}. However, this evolution can be disclosed, if exists, using an approach proposed by \citet{Goobar2002} and \citet{Riess2004b}. The authors estimated extinction produced by intergalactic dust by fitting a theoretical extinction curve to observations of SNe dimming in the redshift interval $z < 0.5$. They showed that a satisfactory fit can be found in this redshift interval, but a remarkable discrepancy appears for higher $z$, see \citet[their fig. 3, model A]{Goobar2002} or \citet[their fig. 7, 'high-$z$ gray dust' model]{Riess2004b}. The discrepancy can be removed when the proper density of intergalactic dust becomes redshift independent for $z > 0.5$ which is not physically well justified. 

However, the approach of \citet{Goobar2002} and \citet{Riess2004b} can be modified to consider the intergalactic extinction more properly. The redshift-dependent optical depth of intergalactic dust in the B-band reads \citep[his eq. 19]{Vavrycuk2017a}
\begin{equation}\label{eq42}
\tau_B\left(z\right) = \int_0^{z} \lambda_B \left(1+z'\right)^2 \,\, \frac{c dz'}{H\left(z'\right)} \,,
\end{equation}
where Hubble parameter $H(z)$ is defined in eq. (19). This formula takes into account an increase of the proper dust density with redshift as $(1+z)^3$ and a decrease of the intergalactic opacity $(1+z)^{-1}$ due to the $1/\lambda$ extinction law.  Consequently, the extinction correction of the distance modulus is expressed as
\begin{equation}\label{eq43}
\Delta m_B \left(z\right) = -2.5 \, log_{10} \, {e^{-\tau_B (z)}} \,.
\end{equation}

Since the extinction correction depends on the Hubble constant $H_0$ and intergalactic extinction $\lambda_B$, we can invert for the optimum values of $H_0$ and $\lambda_B$ by fitting the dust model in the whole interval of the SNe observations. In this way, using the SNe Union2.1 compilation \citep{Suzuki2012}  for the model C in Table 1, we get the best fit for parameters $H_0 = 66.8 \,\, \mathrm{km \, s^{-1} \, Mpc^{-1}}$ and $\lambda_B = 0.094 \,\, \mathrm{Gpc^{-1}}$ (see Fig.~\ref{fig:9}). Analogously, the same data can be inverted for the optimum values of $H_0$ and $\Omega_\Lambda$ describing the flat $\Lambda$CDM model of the accelerating expansion with the best fit for $H_0 = 71.4 \,\, \mathrm{km \, s^{-1} \, Mpc^{-1}}$, $\Omega_\Lambda = 0.76$ and $\Omega_m = 0.24$ (see Fig.~\ref{fig:10}). The both optimum solutions yield comparable normalized misfits: 0.19 (Fig.~\ref{fig:11}b, the black line) for the model of the opaque universe, and 0.18 for the model of the accelerating expansion (Fig.~\ref{fig:11}a, the red line) when the intergalactic extinction is neglected. As illustrated in Fig.~\ref{fig:12}, also residua of individual SNe as a function of redshift are comparable for the both alternative models. Hence, the model of the opaque universe can successfully explain the observed diming of SNe with no need to introduce the dark energy and the accelerating expansion.

\section{Other supporting evidence}

The cyclic cosmological model of the opaque universe successfully explains also other observations. First, the model predicts existence of very old mature galaxies at high redshifts. The existence of mature galaxies in the early Universe was confirmed, for example, by \citet{Watson2015} who analyzed observations of the Atacama Large Millimetre Array (ALMA) and revealed a galaxy at $z > 7$ highly evolved with a large stellar mass and heavily enriched in dust. Similarly, \citet{Laporte2017} analyzed a galaxy at $z \sim 8$ with a stellar mass of $\sim 2 \times 10^9 \,\, M_\odot$ and a dust mass of $\sim 6 \times 10^6 \,\, M_\odot$. A large amount of dust is reported by \citet{Venemans2017} for a quasar at $z = 7.5$ in the interstellar medium of its host galaxy. In addition, a remarkably bright galaxy at $z \sim 11$ was found by \citet{Oesch2016} and a significant increase in the number of galaxies for $8.5 < z < 12$ was reported by \citet{Ellis2013}. Note that the number of papers reporting discoveries of galaxies at $z \sim 10$ or higher is growing rapidly \citep{Hashimoto2018, Hoag2018, Oesch2018, Salmon2018}.

Second, the model is capable to explain the origin of the CMB, its temperature and the CMB anisotropies \citep{Vavrycuk2018}. The opacity of the Universe is mostly caused by absorption of light by intergalactic dust.  The absorbed energy warms up dust grains and produces their thermal radiation in the form of the CMB. Since the intensity of light in intergalactic space is low, the intergalactic dust is cold and emits radiation at microwave wavelengths. The temperature of the CMB is predicted by this theory with the accuracy of 2\%, being controlled by the EBL intensity and by the ratio of galactic and intergalactic opacities. The temperature of intergalactic dust increases linearly with redshift and exactly compensates the change of wavelengths due to redshift. Consequently, dust radiation looks apparently like the blackbody radiation with a single temperature. 

As shown by \citet{Vavrycuk2018}, the CMB power spectrum fluctuations come from the EBL fluctuations, which are produced by large-scale structures in the Universe. The prediction of a close connection between the CMB fluctuations and the large-scale structures is common to the opaque universe model and the Big Bang theory. The arguments of both theories explaining this connection are, however, reversed. The Big Bang theory claims that the large-scale structures are a consequence of the CMB fluctuations originated at redshifts $z \sim  1000$, while the dust theory claims that the CMB fluctuations are produced by the large-scale structures at redshifts less than $3 - 5$. The polarization anomalies of the CMB correlated with temperature anisotropies are caused by the polarized thermal emission of needle-shaped conducting dust grains aligned by large-scale magnetic fields around clusters and voids. The phenomenon is analogous to the polarized interstellar dust emission in our Galaxy, which is observed at shorter wavelengths because the temperature of the galactic dust is higher than that of the intergalactic dust \citep{Lazarian2002, Gold2011, Ichiki2014, Ade2015, Aghanim2016b}.

Third, the model of the opaque universe explains satisfactorily: (1) the observed bolometric intensity of the EBL with a value of $100 - 200 \,\, \mathrm{nW \, m^{-1} \, sr^{-1}}$ \citep{Vavrycuk2017a}, (2) the redshift evolution of the comoving UV luminosity density with extremely high values at redshifts $2 < z < 4$ \citep[his fig. 11]{Vavrycuk2018}, and (3) a strong decay of the global stellar mass density at high redshifts \citep[his fig. 12]{Vavrycuk2018}. The increase of the luminosity density at $z \sim 2 - 3$ does not originate in the evolution of the star formation rate as commonly assumed but in the change of the proper volume of the Universe. The decrease of the luminosity density at high $z$ originates in the opacity of the high-redshift universe. Importantly, the proposed theory is capable to predict quantitatively all these phenomena by considering a fixed value of the intergalactic opacity, obtained from independent measurements \citep{Xie2015}. 

If the observations are corrected for the intergalactic opacity, the comoving number density of galaxies, the global stellar mass, and the overall dust masses within galaxies and in the intergalactic space are constant with cosmic time \citep[his figs 11 and 12]{Vavrycuk2018}. This property of the universe is essential also for deriving the equations of cosmic dynamics in Section 2 Theory.

\section{Discussion and conclusions}

Physicists and astronomers sought forces counterbalancing the gravity produced by galaxies in the Universe for more than one century. This puzzle was also addressed by \citet{Einstein1917} who introduced a cosmological constant into his field equations of general relativity to establish a pressure acting against the gravity and maintaining a static universe. After a discovery of the expansion of the Universe by \citet{Lemaitre1927} and \citet{Hubble1929}, Einstein abandoned this concept \citep{Einstein_deSitter1932, O'Raifeartaigh_McCann2014}. However, this concept was later revived in the form of dark energy to explain the accelerating expansion reported by teams conducting and interpreting supernovae measurements \citep{Riess1998, Perlmutter1999}. The dark energy concept and the $\Lambda$CDM cosmological model are now widely accepted even though they are speculative with no clear physical basis. 

Surprisingly, the radiation pressure as a cosmological force acting against the gravity has not been proposed yet, even though its role is well known in theory of stellar dynamics \citep{Kippenhahn2012}. The radiation pressure is important in evolution of massive stars \citep{Zinnecker_Yorke2007}, in supernovae stellar winds and in galactic wind dynamics \citep{Martin2005, Hopkins2012}. However, the cosmological consequences of the radiation pressure have been overlooked. This was caused probably by little knowledge of galactic and intergalactic opacities and by unknown behavior of the EBL with redshift. Since the high-redshift universe is commonly assumed to be dark instead of to be opaque, any role of the radiation pressure is automatically excluded at high redshift. By contrast, the radiation pressure plays an essential role in the opaque universe model, because the intergalactic opacity and the intensity of the EBL steeply increase with redshift. This causes that the radiation pressure, negligible at present, was significant at high redshifts and could fully eliminate gravity effects and stop the universe contraction.

In this way, the expansion/contraction evolution of the Universe seems to be a result of an imbalance of gravitational forces and radiation pressure. Since the comoving global stellar and dust masses are basically independent of time with minor fluctuations only, the evolution of the Universe must be stationary. This implies a model of the cyclic universe with an overcritical mass density. Obviously, the recycling processes of stars and galaxies \citep{Segers2016, Angles-Alcazar2017} play a more important role in this model than in the Big Bang theory.  

The age of the Universe in the cyclic model becomes unconstrained and galaxies can be observed at any redshift less than the maximum redshift $z_\mathrm{max}$. The only limitation is high intergalactic opacity, which can prevent observations of the most distant galaxies. Hypothetically, it is possible to observe galaxies from the previous cycle/cycles, if their distance is higher than that corresponding to $z_\mathrm{max}$, which is predicted to be less than $20 - 45$. The identification of galaxies from the previous cycles will be, however, difficult, because their redshift will be a periodic function with increasing distance. We cannot even exclude observations of very distant galaxies characterized by blueshifts, if the rest-frame galaxy emission corresponds to a cosmic epoch characterized by a larger volume of the Universe than at present. However, this is not very likely, because the intergalactic opacity increases with frequency and the radiated light will be strongly attenuated. On the contrary, it is plausible to assume that the observed CMB comes partly from previous cycle/cycles. The intergalactic opacity at the CMB wavelengths is lower by several orders than the opacity at the optical spectrum, hence the CMB traverses enormous distances unless it is fully attenuated.

In the presented theory and in the analysis of the SNe measurements, I simplified the problem and ignored the issue of the spatial curvature of the Universe, which is related to the density of the Universe in general relativity. I developed just a Newtonian form of the equations because of the following reasons. First, except for the curvature of the Universe significantly deviating from the flat space, this issue is rather minor. Second, the problem of the spatial curvature due to an extremely inhomogeneous density distribution in the Universe is far more involved than usually treated. Since the Einstein's equations are non-linear, inserting simple large-scale averages into the equations might be incorrect and can yield biased results. This problem was addressed by \citet{Wiltshire2009}, \citet{Duley2013} and \citet{Dam2017}, who developed the so-called timescape cosmology as an alternative to the homogeneous cosmology represented by, e.g., $\Lambda$CDM cosmological model. Obviously, refining the presented theory by including a correctly treated spatial curvature is desirable in future research.

In summary, the opaque universe model and the Big Bang theory are completely different concepts of the Universe. The both theories successfully predict basic astronomical observations such as the Universe expansion, the luminosity density evolution with redshift, the global stellar mass history, SNe measurements and the CMB observations. However, the Big Bang theory is based on many speculative assumptions and hypotheses supported by no firm evidence. This applies to speculations about an initial singularity, inflation, Big Bang nucleosynthesis, baryon acoustic oscillations, relic radiation, reionization epoch and accelerating expansion, which are full of inconsistencies, anomalies or tensions \citep{Buchert2016, Bullock2017}. In addition, unphysical concepts as the non-baryonic dark matter and the dark energy are introduced to reconcile the predictions with observations \citep{Bull2016}. By contrast, the cyclic model of the opaque universe is based on the standard physics, it is less speculative and predicts better the current observations. The dark matter revealed by rotation curves of spiral galaxies and gravity measurements \citep{Rubin_Ford1970, Rubin1980, Bahcall1995, Bahcall_Kulier2014, Bahcall2015} plays a key role in this theory, but there is no need to assume its non-baryonic origin. The existence of the non-baryonic dark matter is questioned also by other studies \citep{Kroupa2015}, e.g., by a detailed study of properties of faint satellite galaxies of the Milky Way \citep{Kroupa2010}, a failure of the dual dwarf galaxy theorem \citep{Kroupa2012}, and by recent observations of a radial acceleration relation of galaxies \citep{McGaugh2016, Lelli2017}.

Further development of the proposed cosmological model and more definite conclusions about its validity are conditioned by more accurate measurements of the spectrum and anisotropies of the EBL and CMB, physical properties of the intergalactic medium and Lyman α systems, mass density and the curvature of the Universe, observations of galactic and intergalactic opacities, extinction laws of interstellar and intergalactic dust, more accurate SNe Ia measurements, and observations of galaxies at redshifts $z > 10$.


\bibliographystyle{aa}
\bibliography{paper} 

\end{document}